%% file: paper.tex
\documentclass[preprint2,numberedappendix]{emulateapj-rtx4}
\usepackage{natbib}
\usepackage{amsmath}
\usepackage{graphicx,times,bm,url}
\usepackage{afterpage,lscape}
\graphicspath{{./fig/}{./png/}}
\include{macros}
\usepackage{color}

\begin{document}

\title{Effects of Fieldline Topology on Energy Propagation in the Corona}

\author{S. Candelaresi}
\author{D. I. Pontin}
\author{G. Hornig}
\affil{
Division of Mathematics, University of Dundee, Dundee, DD1 4HN, UK
}

\begin{abstract}
We study the effect of photospheric footpoint motions on magnetic
field structures containing magnetic nulls.
The footpoint motions are prescribed on the photospheric boundary as a velocity field which entangles the magnetic field.
We investigate the propagation of the injected energy,
the conversion of energy, emergence of current layers
and other consequences of the non-trivial magnetic field topology in this situation.
These boundary motions lead initially to an increase in magnetic and kinetic energy. Following this,
the energy input from the photosphere is partially
dissipated and partially transported out of the domain through the Poynting flux.
The presence of separatrix layers and magnetic null-points fundamentally alters the
propagation behavior of disturbances from the photosphere into the corona.
Depending on the field line topology close to the photosphere, the energy is either
trapped or free to propagate into the corona.
\end{abstract}

\section{Introduction}
From observations
and field extrapolations \cite[e.g.][]{Longcope-Brown-2003-10-8-PhysPlasm,platten2014}
we know that the solar magnetic field has a rather complex structure.
Apart from its solar-scale toroidal and poloidal field, which
is rather weak compared to small-scale contributions, there are
large-scale magnetic field lines connecting back to the photosphere
\cite[e.g.][]{Filippov-1999-185-2-SolPhys} which are anchored at magnetic footpoints.
Such large-scale loops are found both inside and outside active regions
\citep{Larmor-1934-469-94-MNRAS, Gosic-Rubio-2014-797-49-ApJ}.

Magnetic field structures exist also on much smaller scales, and 
we now know that the lower corona is characterized by a so-called
magnetic carpet structure of many short, differently oriented loops
due to mixed polarities of opposite signs over a broad range of 
scales \cite[e.g.][]{schrijver1998}.
Such fields contain a large number of magnetic null points with a
decreasing population density with height
\citep{Longcope-Brown-2003-10-8-PhysPlasm, Edwards-Parnell-2015-SolPhys}.
The presence of these null points and the wide range of field line 
topologies in general -- from short low-lying loops
to long loops that extend high into the atmosphere and
open field lines --
are essential in understanding the propagation of energy
from footpoint motions \cite[e.g.][]{Filippov-1999-185-2-SolPhys, schrijver2010, Santamaria-Khomenko-2015-577-A70-AA}
to the upper layers of the solar atmosphere.

It is now well established that various geometrical or topological features
of the coronal magnetic field are preferential sites for current accumulation
and magnetic reconnection \citep{Lau-Finn-1990-350-672-ApJ,
Bogdanov-Burilina-1994J-59-537-ETPL, demoulin1997,pontincraig2005, aulanier2005, Pontin2007, Effenberger-Craig-2016-SolPhys}.
Such features include magnetic null points and their associated
separatrix surfaces, separator lines (intersections of these separatrix
surfaces), and quasi-separatrix layers 
\citep[see][and references therein]{pontin2011}.
Together the magnetic null points and associated separatrix surfaces
and separators are termed the `magnetic skeleton' of the field.
\cite{priest2002} have proposed that reconnection
at these structures within the Sun's so-called magnetic carpet
could provide an integral contribution to the heating
of the coronal plasma.

In this paper we study the evolution of a coronal magnetic field of
non-trivial topology under the influence of prescribed photospheric
motions. There exist a number of previous studies dealing with such a 
scenario, following two main approaches. In the first, the full coronal
system is simulated, and the overall behavior of the system is 
analyzed - such an approach has been successful in demonstrating 
heating of the coronal plasma for numerically accessible parameter
regimes \cite[e.g.][]{gudiksen2005,bingert2011}. 
The second approach involves using a much simpler model for the coronal 
field and plasma, but has the advantage that the detailed time evolution
of the coronal field structure and energy distribution may be followed.
Previous studies of this nature have focused on configurations in which 
the opposite magnetic polarities on the photosphere are well separated
\cite[e.g.][]{galsgaard2000,mellor2005,demoortel2006}, 
and have demonstrated that reconnection and plasma heating take place. 
By contrast to these studies, here we consider the case in which the photospheric
polarities are inter-mixed -- as observed over a large portion
of the photosphere -- leading to configurations with magnetic
nulls within the coronal volume.

In this work we investigate the effect of footpoint motions on the coronal
magnetic field, in particular the propagation of energy and change in topology.
Throughout this paper we refer to magnetic topology with respect to a reference field,
as in the definition of the relative magnetic helicity.
This implies that two fields only have equivalent topology if one can be transformed into
the other by a smooth deformation that leaves the boundaries undisturbed.
Therefore, the topology is not only distinguished by the distribution of magnetic
null points and separatrix layers, but also by magnetic field line braiding.
Such braiding represents a non-trivial topology of the field since the field lines can
only be `unbraided' by either performing motions on the boundary or allowing
reconnection of field lines in the volume.
It is expected that the topology plays a crucial role in the energy transport.
We present three distinct initial fields and discuss their differences and similarities.
Finally, we conclude with drawing connections to the solar magnetic field.

\section{Model and Methods}
It is expected that the formation of electric current concentrations and
propagation of energy in response to footpoint motions will vary greatly
depending on the magnetic field topology. 
Therefore, we examine three magnetic field configurations as described in
section \ref{sec: setups}, while the fluid is driven using a prescribed
driver at the lower $z$-boundary (see section \ref{sec: boundary driver}).

\subsection{Setups} \label{sec: setups}
The initial magnetic field for all of the simulations is potential.
Three different initial conditions are considered here.
The first we use as a `control' case, and 
simply consists of a homogeneous field in the $z$-direction,
while the others contain magnetic null points and are refered to as magnetic
carpet structures.
They are constructed by positioning magnetic dipoles outside the
physical domain.
The field configurations are chosen such that some field lines close back to the
lower boundary, hence creating a magnetic carpet-like structure. The three different
initial conditions considered are described in turn below.

To simplify the setups we choose an initially homogeneous density of value $\rho_0 = 1$
for all test cases and set the sound speed to $\cs = 1$.
Since the magnetic field strength varies in space, the Alfv\'en speed changes as
well with $v_{\rm A} = |\BB|/\sqrt{\mu_0\rho}$, with the magnetic field
$\BB$ and the vacuum permeability $\mu_0$, which we set to $1$.

\subsubsection{Homogeneous Field}
The homogeneous magnetic field is simply given by
\EQ
\BB = B_0\eee_z,
\EN
where we choose $B_0 = 0.25$.
Since the MHD code we apply for our simulations uses the magnetic vector potential we need
to express $\BB$ in terms of the magnetic vector potential $\AAA$ with $\BB = \nab\times\AAA$:
\EQ \label{eq: B hom}
\AAA = \frac{1}{2}
B_0
\left(
\begin{array}{c}
-y \\
x \\
0
\end{array}
\right).
\EN
For this configuration the domain is chosen to be $-4 \le x \le 4$,
$-4 \le y \le 4$ and $0 \le z \le 48$
with a spatial resolution of $256^3$ grid points.

\subsubsection{Embedded Parasitic Polarities}
\begin{figure*}[t!]\begin{center}
\includegraphics[width=2\columnwidth]{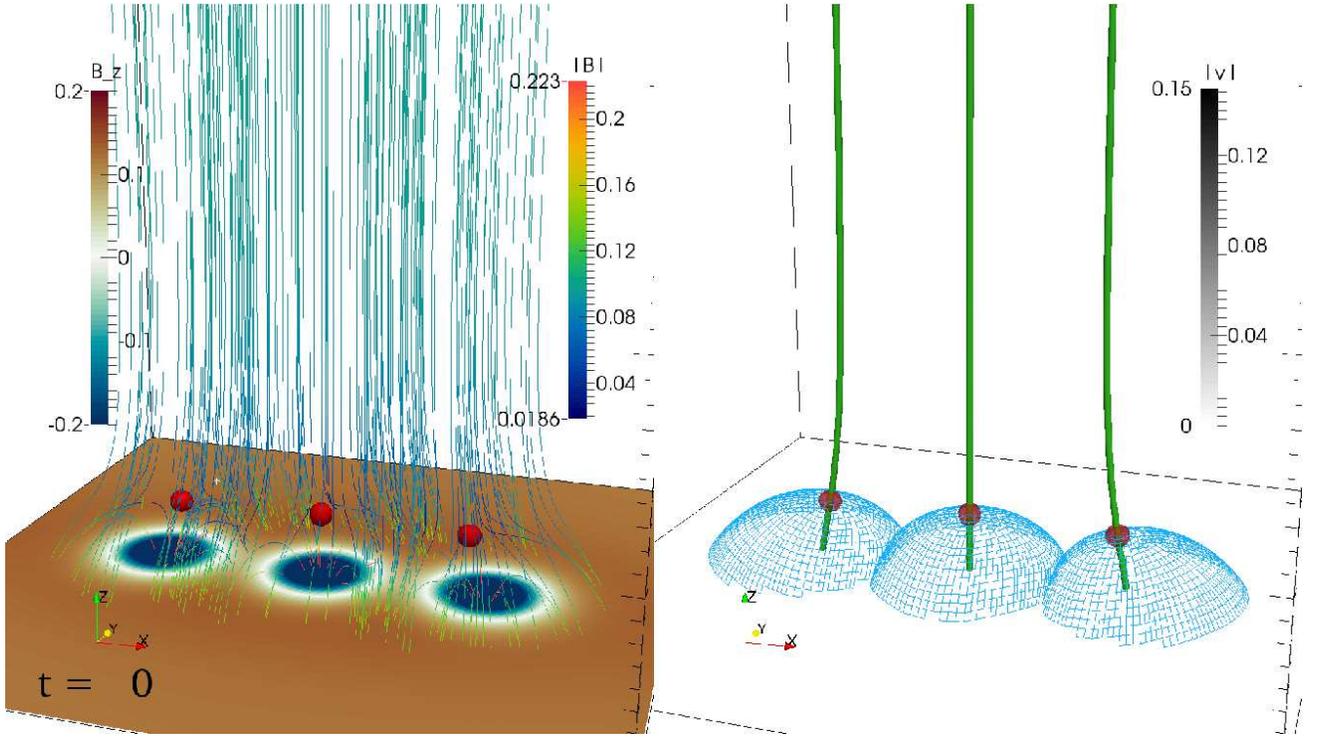}
\end{center}
\caption[]{
Initial condition for the embedded parasitic polarities.
The field lines on the left panel are tracing the magnetic field where the
color denotes the strength of the magnetic field, the red spheres
mark the locations of the magnetic nulls
and the color at the lower boundary denotes
the $z$-component of the magnetic field.
The right panel also shows the magnetic nulls together with the separatrix surfaces
as blue wire frame and the magnetic spines as green tubes.
(The time evolution of this configuration is available as an animation.)
}
\label{fig: parasitic_b0}
\end{figure*}

In the first mixed polarity case considered the photosphere consists of magnetic flux concentrations
embedded within a weaker uniform polarity region of the opposite sign, such that the total flux of the uniform
polarity dominates.
Therefore, in this case the field at large distances along the loop has the same
sign as this uniform background field, while the embedded photospheric magnetic
flux concentrations of opposite sign constitute `parasitic polarity' regions.
Above each of these parasitic polarities is a separatrix dome that encloses
all of the flux that connects from the parasitic polarity back to the
photosphere -- distinguishing it from flux that connects from the photosphere
up to the body of the loop (and the top boundary).
Some sample magnetic field lines are plotted in \Fig{fig: parasitic_b0},
together with the magnetic skeleton that includes the separatrix surfaces.

The magnetic field is constructed by placing three magnetic dipoles at locations
outside our domain of interest.
Specifically, we restrict our studies to the half-space $z>0$, where $z=0$
represents the photosphere, and place all dipoles at $z<0$.
The vector potential for this magnetic field is given by 
\begin{equation}\label{eq: dipoleb}
\AAA=\AAA_0 + B_0\sum_{i=1}^n \epsilon_i \frac{\ee_z\times(\xx-\xx_i)}{|\xx-\xx_i|^3},
\end{equation}
where $\AAA_0$ is the background magnetic field from equation \eqref{eq: B hom},
$\xx_i$ are the locations and $\epsilon_i$ are the strengths of the dipoles.
Here we take $n=3$, $\epsilon_{1,2,3}=-2$, $\xx_1=(0,0,-0.85)$, $\xx_2=(2,0,-0.85)$,
$\xx_3=(-2,0,-0.85)$ and $B_0 = 0.1$.
To make the field quasi-periodic at the $x$ and $y$ boundaries 
(and thus ensure that the field lines within the loop are approximately 
tangent to these boundaries) 
we also add
mirror dipoles in the 8 squares surrounding the computational domain in $x$ and $y$, also at $z=-0.85$.
The domain extends for this configuration to
$-4 \le x \le 4$, $-4 \le y \le 4$ and $0 \le z \le 16$
with a spatial resolution of $256\times256\times512$ grid points.

\subsubsection{Embedded Dominant Polarities}
\begin{figure*}[t!]\begin{center}
\includegraphics[width=2\columnwidth]{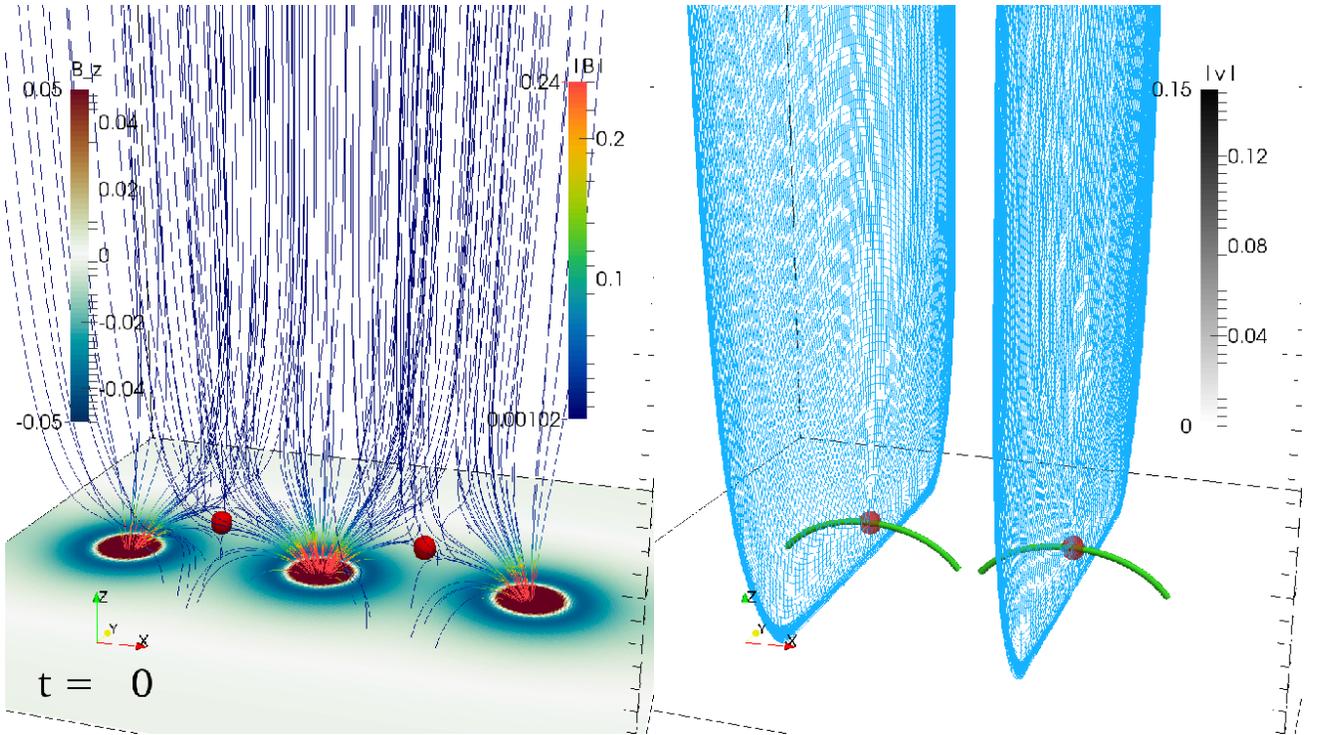}
\end{center}\caption[]{
Initial condition for the embedded dominant polarities.
The field lines on the left panel are tracing the magnetic field where the
color denotes the strength of the magnetic field, the red spheres
mark the locations of the magnetic nulls
and the color at the lower boundary denotes
the $z$-component of the magnetic field.
The right panel also shows the magnetic nulls together with the separatrix surfaces
as blue wire frame and the magnetic spines as green tubes.
(The time evolution of this configuration is available as an animation.)
}
\label{fig: dominant_b0}
\end{figure*}

As a contrast to the above parasitic polarity case we also run simulations in which the
embedded localized polarity regions form the flux of the loop (requiring that the total
flux through the photosphere in our domain of interest is dominated by these polarities).
We refer to this case as embedded dominant polarities.
As shown in \Fig{fig: dominant_b0}, this results in the field lines taking on the
classic `wine glass' shape. 
As shown in the right-hand frame of \Fig{fig: dominant_b0},
the magnetic field in this case also contains magnetic null points, but in this case the
associated separatrix surfaces do not close over regions of the photosphere, but rather
extend vertically along the length of the loop, separating the flux associated with each
embedded dominant polarity, in a manner reminiscent of the {\it coronal tectonics} model
of \cite{priest2002}.

The magnetic field setup that we use is again defined by equation (\ref{eq: dipoleb}),
this time with parameters as follows: $n=3$, $\epsilon_{1,2,3}=0.3$, $\xx_1=(0,0,-0.3)$,
$\xx_2=(2.5,0,-0.3)$, $\xx_3=(-2.5,0,-0.3)$ and $B_0 = 0.1$.
Similarly to the parasitic polarity setup we place mirror dipoles
below the 8 squares surrounding the computational domain in the $xy$-plane.
We also choose the same domain size as previously, specifically
$-4 \le x \le 4$, $-4 \le y \le 4$ and $0 \le z \le 16$
with a spatial resolution of $240\times240\times480$ grid points.

\subsection{Numerical setup}
In order to capture the full effects of magnetic diffusion and reconnection
we solve for the evolution of our
magnetized fluid under the full magnetohydrodynamics (MHD) equations for a resistive,
viscous, isothermal, and compressible gas
\EQ
\frac{\partial \AAA}{\partial t} = \uu \times \BB + \eta\nabla^{2}\AAA,
\label{eq: dAdt}
\EN
\EQ
\frac{\DD \uu}{\DD t} = -\cs^{2} \nab \ln{\rho} +
\JJ\times\BB/\rho + \FF_{\rm visc},
\label{eq: dUdt}
\EN
\EQ
\frac{\DD \ln{\rho}}{\DD t} = -\nab \cdot \uu,
\label{eq: drhodt}
\EN
with the magnetic vector potential $\AAA$, velocity $\uu$,
magnetic field $\BB = \nab\times\AAA$, magnetic resistivity $\eta$,
isothermal speed of sound $\cs$, density $\rho$,
current density $\JJ = \nab\times\BB$,
viscous forces $\FF_{\rm visc}$
and Lagrangian time derivative
$\DD/\DD t = \partial/\partial t + \uu\cdot\nab$.
Here the viscous forces are given as $\FF_{\rm visc} = \rho^{-1}\nab\cdot2\nu\rho\SSS$,
with the kinematic viscosity $\nu$, and traceless rate of strain tensor
$\SSS_{ij} = \frac{1}{2}(u_{i,j}+u_{j,i}) - \frac{1}{3}\delta_{ij}\nab\cdot\uu$.
Being an isothermal gas we have $p = \cs^{2}\rho$ for the pressure.
For the vector potential $\AAA$ we apply the Weyl gauge with $\nab\cdot\AAA = 0$.

Throughout our simulations we use $\eta = 4\times10^{-4}$ to reduce
magnetic energy dissipation as much as the limited resolution allows.
For the kinematic viscosity we choose $\nu = 10^{-4}$ for the homogeneous
initial field and $4\times 10^{-3}$ for the other simulations.
This is necessary to dissipate the stresses that build up in the 
vicinity of the lower boundary.

\Eqss{eq: dAdt}{eq: drhodt} are solved using the {\sc Pencil Code}
which is an Eulerian finite difference code using sixth-order in space derivatives
and a third-order time stepping scheme \citep{BD02PC}.

\subsection{Boundary Conditions}
Any flow through the side boundaries ($x$ and $y$) and the lower
boundary ($z_0$) is suppressed, as we
set the normal component of the velocity field to zero, while the
tangential component is free (derivative across the boundary is zero).
For the upper boundary the velocity can, in principle, reach any
value, as we set all components antisymmetric with respect to the
boundary value.
On the lower boundary a tangential flow is 
prescribed, using the method described in Section \ref{sec: boundary driver}.

The boundary conditions for the density are set to symmetric at all boundaries, which forces its
derivative across the boundaries to zero, but does not directly restrict its
value.
With the isothermal equation of state, this implies that the pressure forces
across the boundaries are zero.

For the magnetic field we set the $x$ and $y$ components of the vector potential
to be antisymmetric with respect to the boundary value at the $x$ and $y$
boundaries, while the $z$-component is symmetric.
This unusual condition is needed due to the presence of a mean magnetic field
in $z$-direction for which the vector potential increases linearly in magnitude
with distance from the projected center.
At the upper boundary we choose all three components of the magnetic vector potential
to be antisymmetric with respect to the boundary value.

For the lower boundary we choose two different conditions, depending on
the initial field, to ensure that any initially potential field is also
potential, i.e.\ current-free, at the boundary.
This is achieved by extrapolating the field into the ghost zones
via a potential field extrapolation.
For the homogeneous initial condition we choose the same conditions at $z_0$
for the magnetic vector potential as for the side boundaries.
While the used extrapolation routine renders the parasitic polarities field to
be potential to a good approximation
at the lower boundary, for the dominant polarities case we observe a small
``residue'' non-potentiality near $z = z_0$.
This has consequences for the field's initial dynamics before the system damps
away those small deviations.

\subsection{Wave Damping Region}
We wish to simulate an upper boundary which is open for Alfv\'enic waves
and analyze the energy propagation into the corona without
the complicating effects of reflection from the opposite loop footpoints.
However, as specified, the upper boundary condition for the magnetic vector potential
is such that Alfv\'enic waves reflect, rather than leave the domain.
This would lead to the interference of the upwards and downwards traveling
waves with possible accumulation of magnetic energy in the domain.
We therefore impose a wave damping region for the embedded and parasitic
polarity configuration which extends from $z = 8$ to the top of the domain
at $z = 16$, in which we increase the viscosity by a factor of $8$
within an interval of length $1$ at $z = 8$ via a step like function
via a hyperbolic tangent variation.
As the reflected damped wave returns from the damping region its intensity
is only a fraction of what it was initially, which is typically less than $7.7\%$
of the amplitude of the wave entering the damping region.
Our subsequent energy dissipation and flux calculations are preformed on the domain
excluding the wave damping region.
We omit the wave damping region for the homogeneous case, since we stop
the simulation as soon as the first disturbance reaches the upper boundary.

\subsection{Energy Dissipation and Fluxes}
In our isothermal compressible system, kinetic energy and magnetic energy
can be transformed into one other through the action of the Lorentz 
force, and in addition each may be dissipated by the resistive 
and viscous terms -- this energy being lost to the system due to the 
isothermal assumption.
Since the boundary conditions allow for magnetic energy fluxes out of the domain we also need to take those
into account when considering the overall energy balance.

\subsubsection{Magnetic Energy}
Starting from the induction equation \eqref{eq: dAdt} we can derive the
form for the magnetic energy variation as
\begin{eqnarray} \label{eq: em}
\frac{\dd}{\dd t} E_{\rm M} & = &
\frac{1}{2} \frac{\dd}{\dd t} \int_{V} {\BB^2}\ \dd V \nonumber \\
 & = & \int_{V} \underbrace{(-\JJ\times\BB)\cdot\uu}_{\rm -Lorentz} -
 \underbrace{\eta\JJ^2}_{\rm ohmic})\ \dd V + \nonumber \\
 & & \int_{\partial V} \underbrace{((\BB\cdot\uu)\BB - \BB^2\uu - \eta\JJ\times\BB)}_{\rm fluxes}\cdot\nn\ \dd S,
\end{eqnarray}
where the first integral is over the domain $V$ which excludes the
wave damping region and the second is a
surface integral over the boundary $\partial V$ with normal vector $\nn$ pointing
outside the domain and $\dd S$ being the infinitesimal surface element on
$\partial V$.

The different terms in equation \eqref{eq: em} are the work done by the
Lorentz force, the ohmic dissipation and the three flux terms at the
boundaries, respectively.
We will consider each of the five terms separately.

\subsubsection{Kinetic Energy}
Similar to the calculations for the magnetic energy we can use the
momentum equation \eqref{eq: dUdt} and the continuity equation
\eqref{eq: drhodt} to compute the
different terms for the kinetic energy flux and dissipation to obtain
\begin{eqnarray} \label{eq: ek}
\frac{\dd}{\dd t} E_{\rm K} & = &
\frac{1}{2} \frac{\dd}{\dd t} \int_{V} \rho {\uu^2}\ \dd V \nonumber \\
 & = & \int_{V} \underbrace{-\cs^2\uu\cdot\nab\rho}_{\rm compression} +
 \underbrace{(\JJ\times\BB)\cdot\uu}_{\rm Lorentz} +
 \underbrace{\rho\uu\cdot\FF_{\rm visc}}_{\rm viscous}\ \dd V + \nonumber \\
 & & \frac{1}{2} \int_{\partial V} \underbrace{-(\uu^2\rho\uu)\cdot}_{\rm fluxes}\nn\ \dd S.
\end{eqnarray}

The terms are the gas compression term through which kinetic energy is
dissipated into heat (and lost from the system),
the work done by the Lorentz
force, which couples the magnetic field with the fluid, the viscous
dissipation and the fluxes through the boundaries, respectively.
As with the magnetic energy, we will consider each of the four terms separately.

\subsection{Boundary Driver} \label{sec: boundary driver}
\begin{figure}[t!]\begin{center}
\includegraphics[width=\columnwidth]{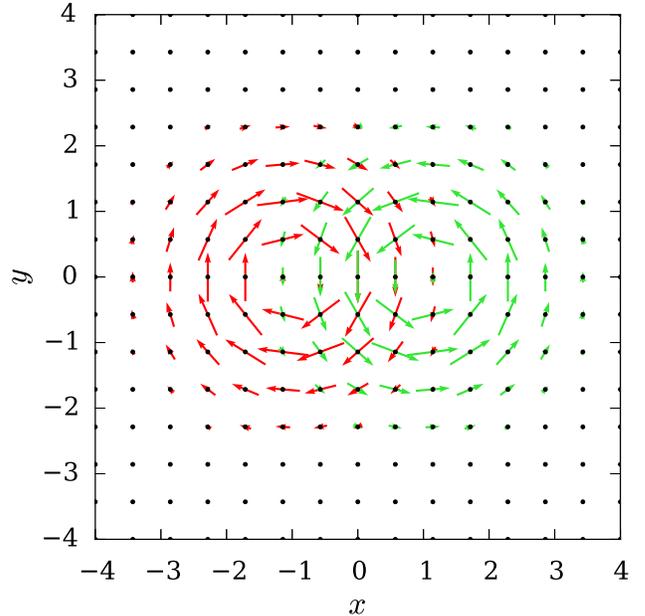}
\end{center}\caption[]{
Representation of the footpoint driving velocity  at two different times in red (left)
and green (right) arrows.
We switch between the two driving vortices every $32$ code units.
}
\label{fig: driver}
\end{figure}

Photospheric foot point motions are simulated by imposing a time and space
varying velocity field at the lower ($z = z_0$) boundary.
Any existing magnetic field which connects to this boundary is then subjected
to this driving.
For the driving velocities we prescribe a blinking vortex pattern,
which when applied on the boundary of an initially homogeneous
magnetic field in an ideal fluid would create the so called $E^3$ braid of
\cite{wilmotsmith2009a}.
The evolution of the homogeneous field
under continued application of such a boundary driving pattern was
recently considered by \cite{Ritchie2016}.
The driving flow consists of two (partially overlapping) circular regions 
at which opposite twisting motions
are applied. The timescale of the driver is such that within the 
time of the simulations a total of six twisting motions are applied
at the photosphere (three of each sign).
Specifically, we force the velocity at $z = z_0$ towards the following profile:
\begin{eqnarray}\label{eq: driver E3}
u_{\rm d}^{x}(x, y, z_0) & = & \pm u_0\exp\left[(-(x\mp x_c)^2-y^2)/2-\right. \nonumber \\
 & & \left.({\rm mod}(t, t_{E3}))^2/(t_{E3}/4)^2\right](-y) \\
u_{\rm d}^{y}(x, y, z_0) & = & \pm u_0k_c\exp\left[(-(x\mp x_c)^2-y^2)/2-\right. \nonumber \\
 & & \left.({\rm mod}(t, t_{E3}))^2/(t_{E3}/4)^2\right](x\mp x_c) \\
u_{\rm d}^{z}(x, y, z_0) & = & 0.
\end{eqnarray}
Here we use $t_{E3} = 32$ and $x_c = 1$.
Our choice of $t_{E3}$ is motivated by the Alfv\'en travel time of $192$ time units for our
box of $48$ in length and Alfv\'en speed of $0.25$, which requires a cadence of $32$ time units
in order to fit 6 twist regions into the domain before the first hits the upper boundary.
The modulo function ${\rm mod}$ is used to simulate the $z$-dependence of the magnetic field.
More precisely, it is given as
\EQ
\pm = \left\{
\begin{array}{c}
+ \quad {\rm if} \quad {\rm mod}({\rm int}(t/t_{E3}), 2) = 0 \\
- \quad {\rm if} \quad {\rm mod}({\rm int}(t/t_{E3}), 2) \ne 0
\end{array}
\right. ,
\EN
with the integer function ${\rm int}$.
In \Fig{fig: driver} we plot a representation of the driver at two different
times with twist injections on the left and right half of the domain.

For our driver we have in mind a setting at the lower part of the corona with lower densities
such that back reactions from the magnetic field to the fluid can be significant.
Furthermore, any direct imposition of the velocity at the lower boundary would create a strong shear
between the boundary and the first inner layer of the simulation box.
Therefore, we force the plasma velocity $\uu$ towards the velocity $\uu_{\rm d}$ at $z = z_0$
through an exponential saturation of the form
\EQ \label{eq: uu exp}
\frac{\partial \uu}{\partial t} =  (\uu - \uu_{\rm d})/\lambda_u,
\EN
with the saturation half time $\lambda_u$. 
We choose $\lambda_u = 0.3$ for the homogeneous case and $\lambda_u = 0.01$
for the other two test cases which ensures a reasonably fast saturation
for the velocity.
Note that due to its nature, the driver can be also counter acted by
forces from the magnetic field.
This back reaction depends on the geometry of the field and can lead
to a non-saturating velocity.

\subsection{Magnetic Skeleton}
We expect the magnetic topology to undergo drastic changes due to the
boundary driver.
The magnetic skeleton, that comprises the stable and unstable manifolds connected
to magnetic null points, characterizes the magnetic topology and separates
the domain into regions of different magnetic connectivity.
Hence the emergence or annihilation of magnetic nulls in the domain indicate
major changes in the magnetic topology. 
Magnetic null points may merge
or be created in pairs -- in each case one null of the pair must have
topological degree $+1$ and the other topological degree $-1$ 
\citep{Fukoka-Ugai-1975-29-133-RISRJ, Greene-1988-93-A8-JGRA,
hornig1996,Murphy-Parnell-2015-22-10-PhysPlasm}.
They can also appear through the boundary, which is essentially open to
magnetic flux.
As a result, separatrix surfaces, which separate areas of different 
magnetic connectivity,
may appear and disappear within the volume during the evolution. 
Since the processes of null pair creation/annihilation
require a non-ideal evolution, they are of interest in understanding 
reconnection and heating in the volume
\citep{wyper2014b,Murphy-Parnell-2015-22-10-PhysPlasm,olshevsky2015}. 
We therefore analyze the 
evolution of the magnetic skeleton during the simulations.

We find the magnetic nulls in our simulations using the trilinear
extrapolation of the magnetic field, which assumes a sufficiently linearizable
field at sub grid scale \citep{Haynes-Parnell-2007-14-8-PhysPlasm}.
To find the separatrix surfaces we use the ring method (which can be found in e.g.\
\cite{Haynes-Parnell-2010-17-9-PhysPlasm}) in which we trace magnetic
field lines from points around the magnetic nulls.
Similarly, we find the spines corresponding to the fan separatrix 
surfaces by tracing out magnetic field lines.

\section{Results}

\subsection{Injection of Braiding}
As a proof of concept we inject the $E^3$ braid
\citep{wilmotsmith2009a} into the initially homogeneous field region using the
prescribed driver (equations \eqref{eq: driver E3}--\eqref{eq: uu exp}) and the parameter
$u_0 = 0.5$.

\begin{figure}[t!]\begin{center}
\includegraphics[width=\columnwidth]{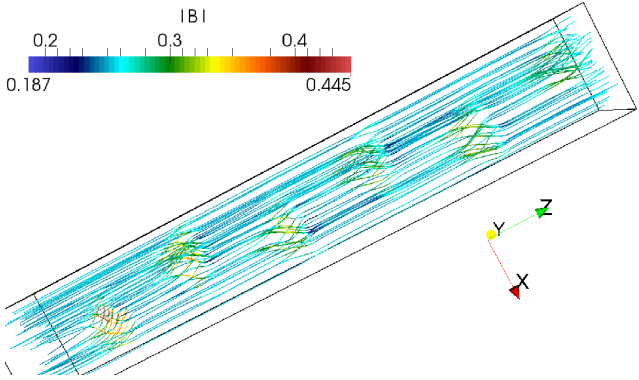}
\end{center}\caption[]{
Magnetic field lines for the initially homogeneous case at $t = 192$ where the
colors denote the field strength $|\BB|$.
}
\label{fig: streamlines hom_t192}
\end{figure}

As we expect, the disturbances from the footpoint motion travel
into the domain via (torsional) Alfv\'en waves.
This leads to a buildup of twisting regions which move into the domain.
As end result we obtain a magnetic field configuration that 
resembles the expected $E^3$ braid (\Fig{fig: streamlines hom_t192}).
This illustrates the efficacy of the footpoint motions to change the
topology of the magnetic field in the case where all field lines are
`open', so that disturbances propagate freely into the domain until they
reach the top boundary.

\begin{figure}[t!]\begin{center}
\includegraphics[width=\columnwidth]{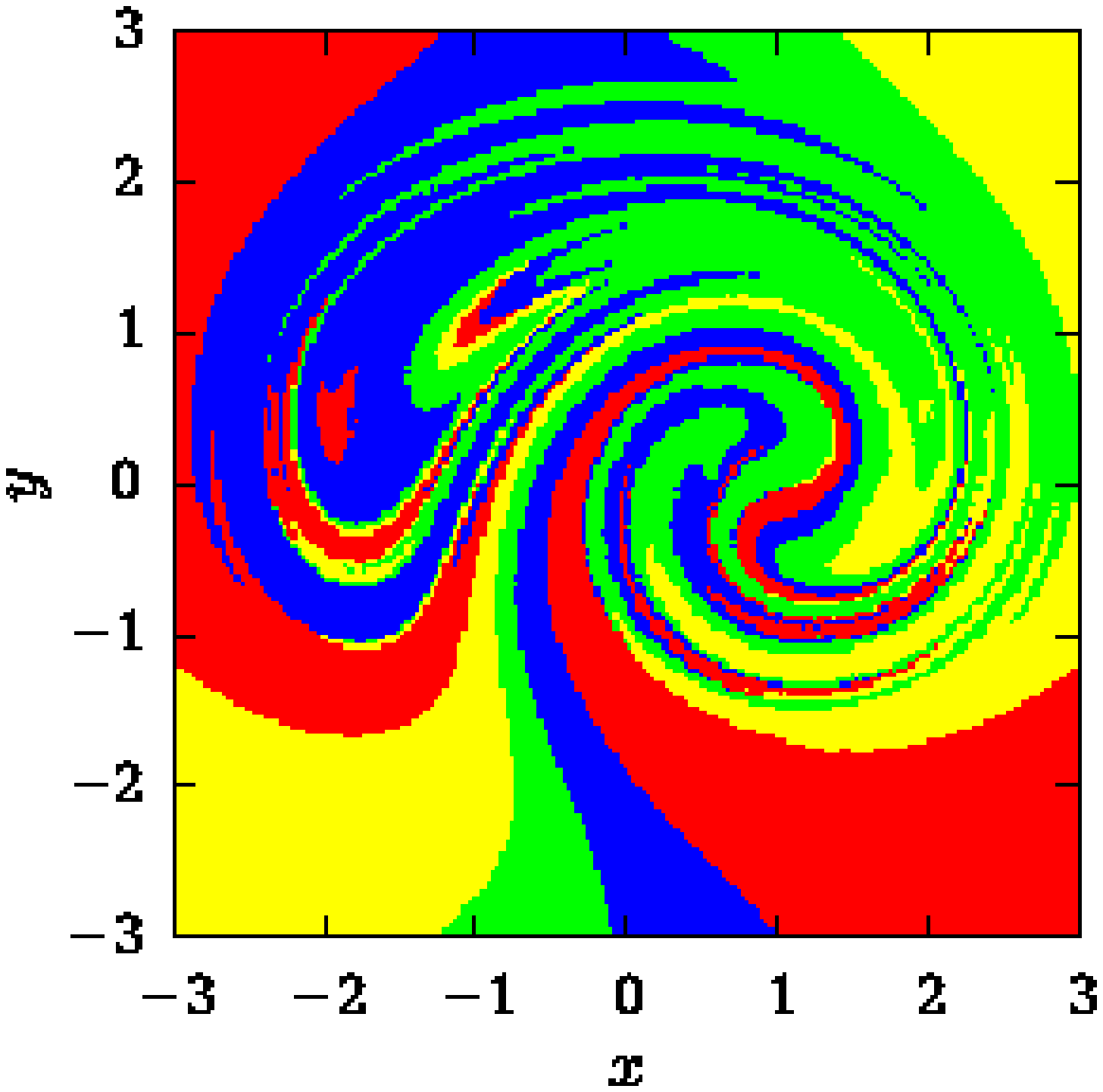} \\
\includegraphics[width=\columnwidth]{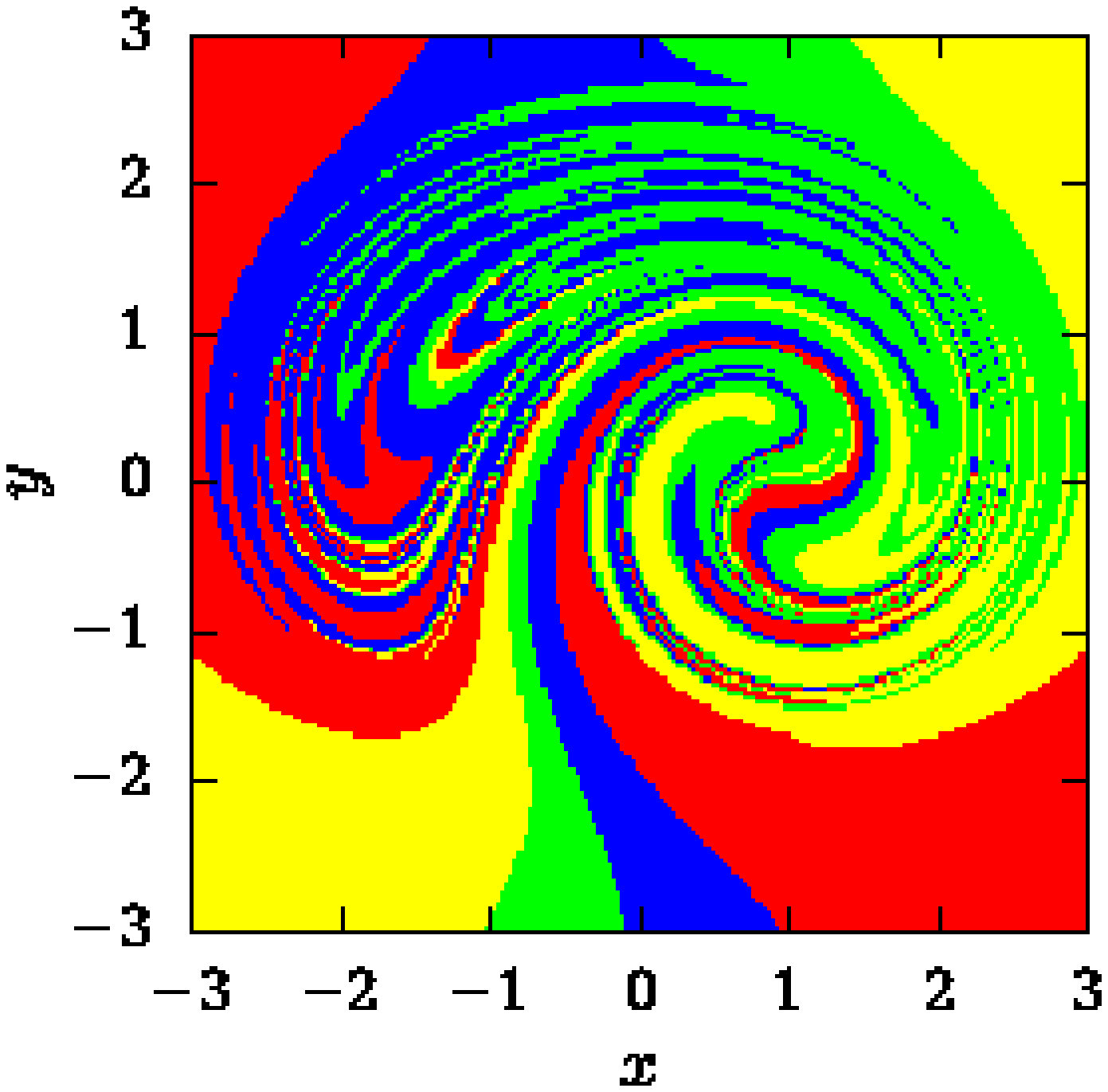}
\end{center}\caption[]{
Color map for the homogeneous field at $t = 192$ (upper panel) and for the
exact $E^3$ field (lower panel).
}
\label{fig: mapping hom_e3}
\end{figure}

Through magnetic resistivity the field is subject to dissipation 
which can lead to small changes of the field line topology even in the absence of intense current layers.
In order to track this we compute the field line mapping given
as the mapping of points $(x, y)$ from the $z = z_0$ plane to the upper boundary
$F(x, y)$ which is induced by the magnetic field lines \citep{Yeates_Topology_2010}.
We then use this mapping to compute the color mapping where we assign
the colors red, blue, green and yellow for $(F_x > x)\wedge (F_y > y)$,
$(F_x > x)\wedge (F_y < y)$, $(F_x < x)\wedge (F_y < y)$ and
$(F_x < x)\wedge (F_y > y)$, respectively (\Fig{fig: mapping hom_e3}).
After time $t = 192$ the field has undergone sufficient braiding to correspond
in the ideal limit to the $E^3$ braid.
We compare the color mapping of our final magnetic field with the color mapping
of the exact $E^3$ field shown in Figure 2 of
\cite{Yeates_Topology_2010}, and find a striking agreement.
The small differences are due to the small but finite magnetic resistivity
which leads to magnetic field dissipation.
The resulting field is then topologically somewhat simpler than the exact $E^3$ braid
presented by \cite{Yeates_Topology_2010}.

\begin{figure*}[t!]\begin{center}
\includegraphics[width=2\columnwidth]{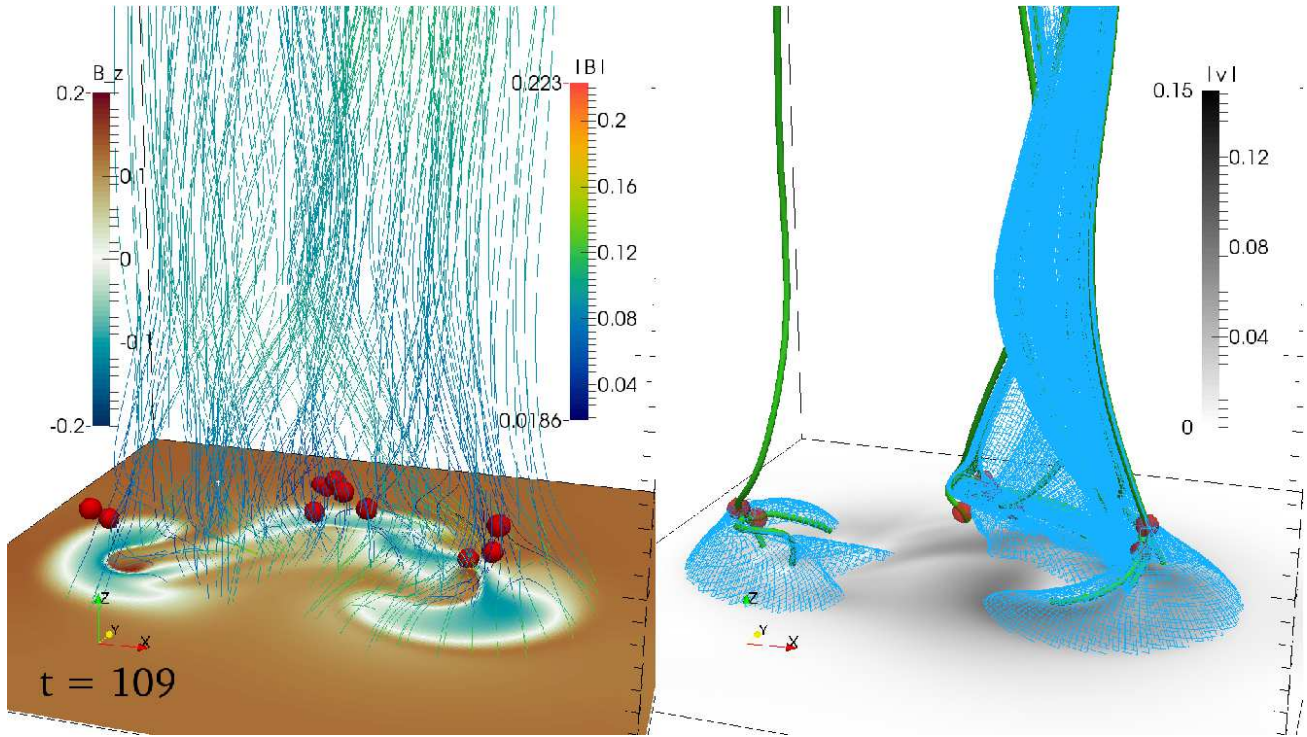}
\end{center}\caption[]{
Magnetic field lines (left) for the parasitic polarities case at $t = 109$ where the
colors denote the field strength $|\BB|$ together with the null points (red spheres)
and the $z$-component of the magnetic field at the lower boundary.
The right panel shows the magnetic skeleton with the null points (red spheres),
magnetic spines (green tubes) and separatrix surfaces (blue grid surface) together with
the magnitude of the velocity at the lower boundary.
(This figure is available as an animation.)
}
\label{fig: skeleton para_t109}
\end{figure*}

For the two magnetic carpet structures with magnetic nulls and separatrix layers,
the propagation of the boundary motions is significantly restricted by the field topology.
Many of the field lines that have footpoints within the twisting regions close back
to the lower boundary rather than extending to the upper boundary.
This leads to Alfv\'enic waves traveling back to the photosphere, and a buildup
of magnetic stresses at low altitudes.
However, as the forcing continues, the magnetic carpet structure is disrupted,
as described in the following section,
reducing the fraction of the driving region covered by closed magnetic
field lines, and thus allowing a propagation of the Alfv\'enic waves to 
higher altitudes (\Fig{fig: skeleton para_t109}).
We quantify this in Section \ref{sec:propagation_to_altitude}, where
we measure the magnetic energy propagation in the different cases.

\subsection{Magnetic Carpet and Field Topology} \label{sec:skeleton}
As the magnetic carpet gets forced from the photosphere, the magnetic
field topology undergoes various changes.
We observe the creation and annihilation of pairs of nulls, and sometimes
also the surfacing of nulls through the photosphere.
While the former has been observed in the past \citep[e.g.][]{Maclean:2009ft, wyper2014b,Murphy-Parnell-2015-22-10-PhysPlasm,olshevsky2015}
the latter is a rather unstudied phenomenon in MHD simulations (though see 
\cite{brown2001} for a magnetic charge topology model).
It turns out that through its particular evolution at the surface,
the field is being restructured in such a way that it gives rise to
additional magnetic nulls and a rather complex structure of the separatrix surfaces
(\Fig{fig: skeleton para_t109}).

As new nulls appear in the domain, the configurations of the separatrix surfaces and
the spines change as well.
Considering first the case of embedded parasitic 
polarities, we observe at an early time (ca.\ $t = 40$ in video1) in the simulation 
a null point appears through the lower boundary,
between the central and right separatrix domes.
That gives rise to a separator pair connecting the new null to
both the central and right null points.
The separatrix surface of this new null point is bounded 
by the spines of the original right and central nulls.
Therefore, part of it extends down to the photospheric boundary while
another part extends up to the top of the box.
Such a structure 
is often called a `separatrix curtain' \citep{Titov:2011gx}, 
and we see many of these appear and
disappear during the evolution (see \Fig{fig: skeleton para_t109} 
and video1) as the null point bifurcations occur.
Apart from such emerging and submerging of single nulls, we also observe the
annihilation and creation of pairs of nulls with opposite sign in 
the weak field region surrounding the original nulls
(see \Fig{fig: skeleton para_t109}), as predicted by
\cite{Greene-1988-93-A8-JGRA, albright1999,wyper2014b}.

\begin{figure}[t!]\begin{center}
\includegraphics[width=\columnwidth]{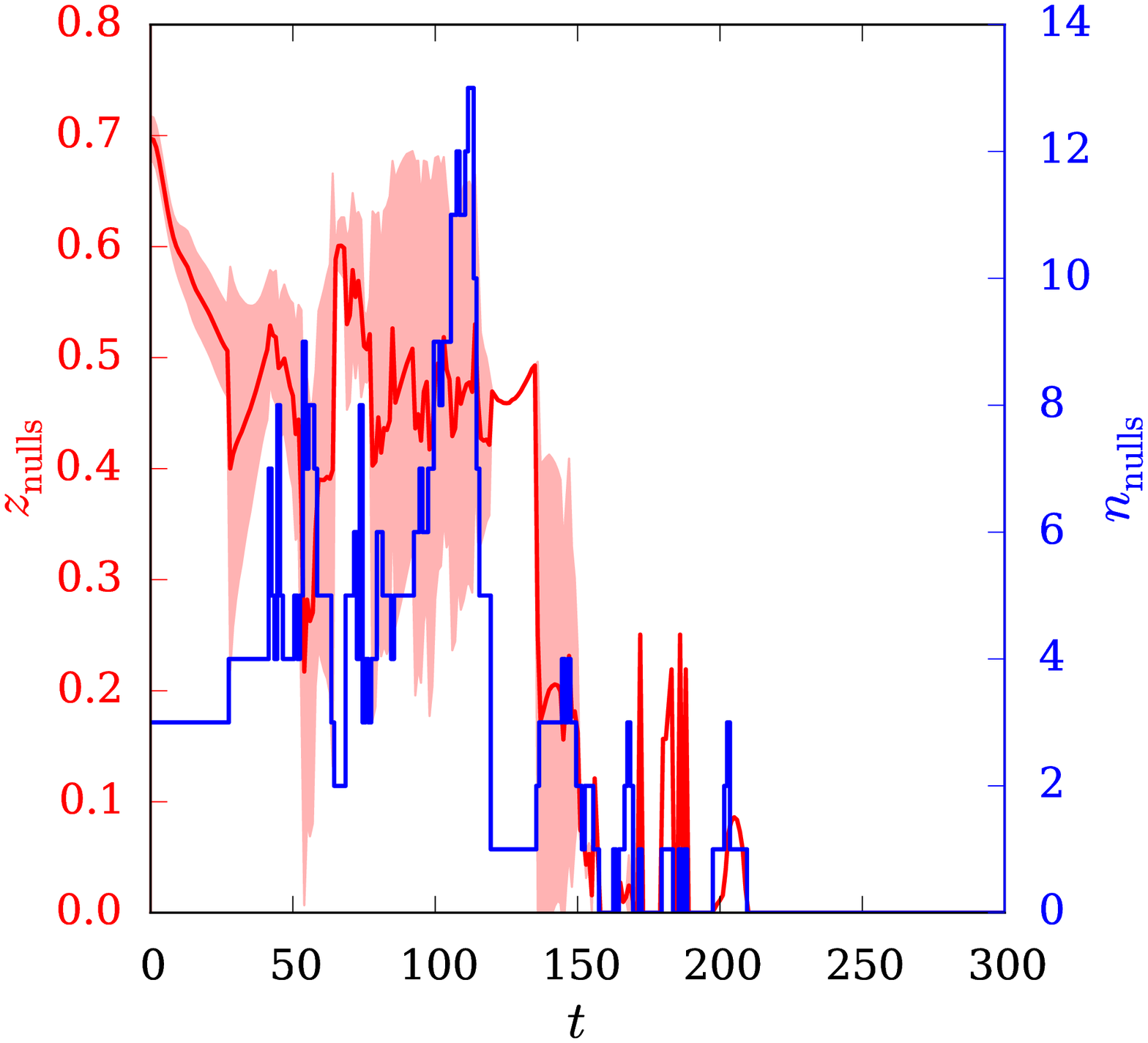} \\
\includegraphics[width=\columnwidth]{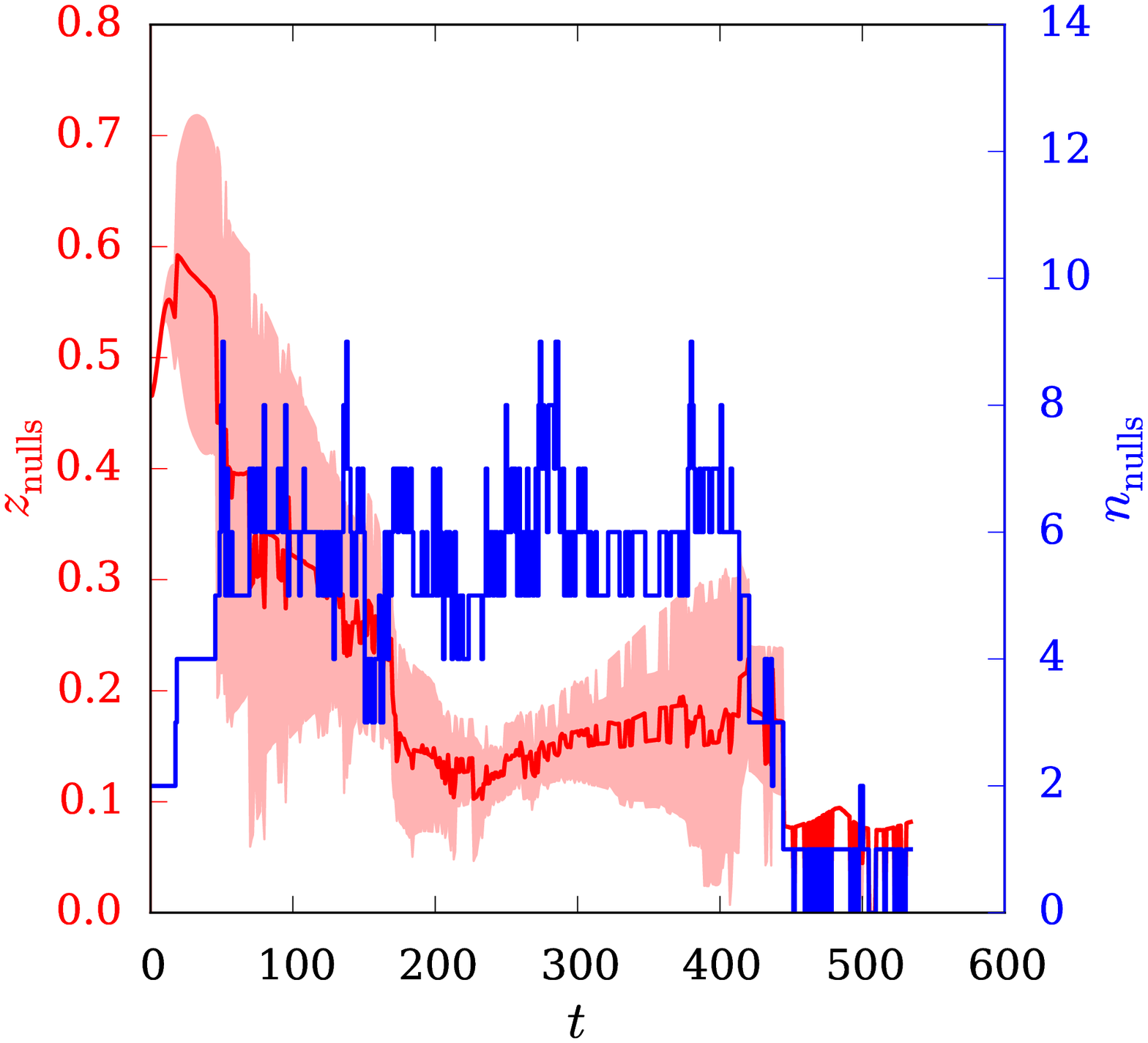}
\end{center}\caption[]{
Average height of the null points $z_{\rm nulls}$ (red line) with standard deviation (light red shading)
together with the total number of magnetic null points $n_{\rm nulls}$ (blue line) in dependence of time
for the parasitic polarity case (upper panel) and dominant polarity case (lower panel).
}
\label{fig: nulls}
\end{figure}

The dynamics of the magnetic null points becomes clearer by computing the average height
of the null points and the number of null points as a function of time (\Fig{fig: nulls}).
The first thing that we observe is that for the parasitic polarity case
the number of null points is much more highly fluctuating, while for the dominant polarity case the null
points are more stable.
For the parasitic polarity case we have bursty production of nulls until $t\approx 100$
(both through topological bifurcations within the domain and null emergence through
the photosphere, as described above).
After this time there is a sharp drop in the average height of the nulls as a result
of the shredding of the polarities, and concurrent with this the null point number drops
rapidly as many nulls leave through the lower boundary.
The number of null points and their average height is intimately linked with
the fraction of the photospheric flux that is `open' to the upper boundary, and thus
has important implications for the propagation of energy and disturbances
from the lower boundary to higher altitudes.
This will be discussed further below.

The above analysis provides a qualitative picture of the propagation of
disturbances into the corona in response to the footpoint motions. In
the following sections we go on to discuss quantitative measures such as 
energy and helicity fluxes.

\subsection{Helicity Injection}
From equation \eqref{eq: em} we know that magnetic energy can be injected from the boundary
as long as the velocity is not perfectly orthogonal to the magnetic field and
$\BB$ is not perpendicular to the surface normal.
For the homogeneous configuration this is the case initially. However, after the first
movement of the footpoints this changes and energy injection is possible and
disturbances of the field propagate through Alfv\'enic waves into the domain.

\begin{figure}[t!]\begin{center}
\includegraphics[width=\columnwidth]{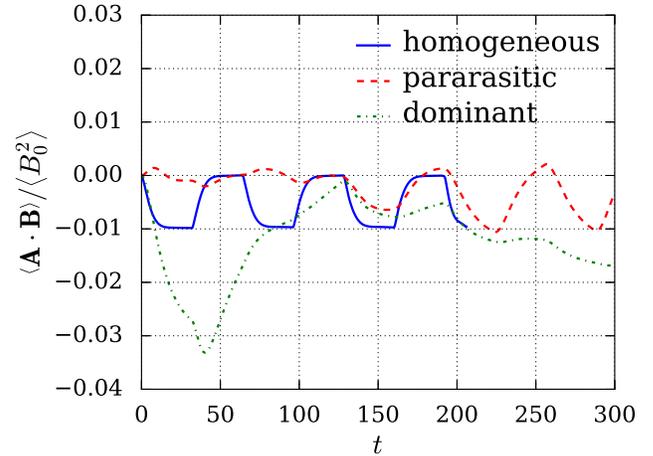}
\end{center}\caption[]{
Time evolution of the normalized magnetic helicity for the three different
configurations.
}
\label{fig: ts all}
\end{figure}

The initially homogeneous field is easily being twisted by the footpoint motions,
which leads to the injection of magnetic helicity for every odd multiple of $t_{E3}$.
We clearly observe this behavior in \Fig{fig: ts all}.
On the other hand, the cases of parasitic and dominant polarities with their
intricate structure and closed (to the photosphere) field lines inhibit
any such propagation initially.
As a consequence, magnetic helicity is not efficiently injected into 
the domain.
However, after sufficient twisting the field realigns itself to a simpler
structure which then allows for efficient propagation of boundary disturbances to large heights.
For that to happen the field needs to reconnect, which is forced by the footpoint
motions.

\subsection{Energy Fluxes, Conversion and Dissipation}
We now calculate the individual contributions to the change in time of the kinetic
and magnetic energy for the embedded parasitic polarity and the dominant
polarity case by applying equations \eqref{eq: em} and \eqref{eq: ek}.
Since the wave damping region at $z > 8$ lies conceptually outside the
physical domain of interest we perform the integrals within $z \le 8$.

\begin{figure}[t!]\begin{center}
\includegraphics[width=\columnwidth]{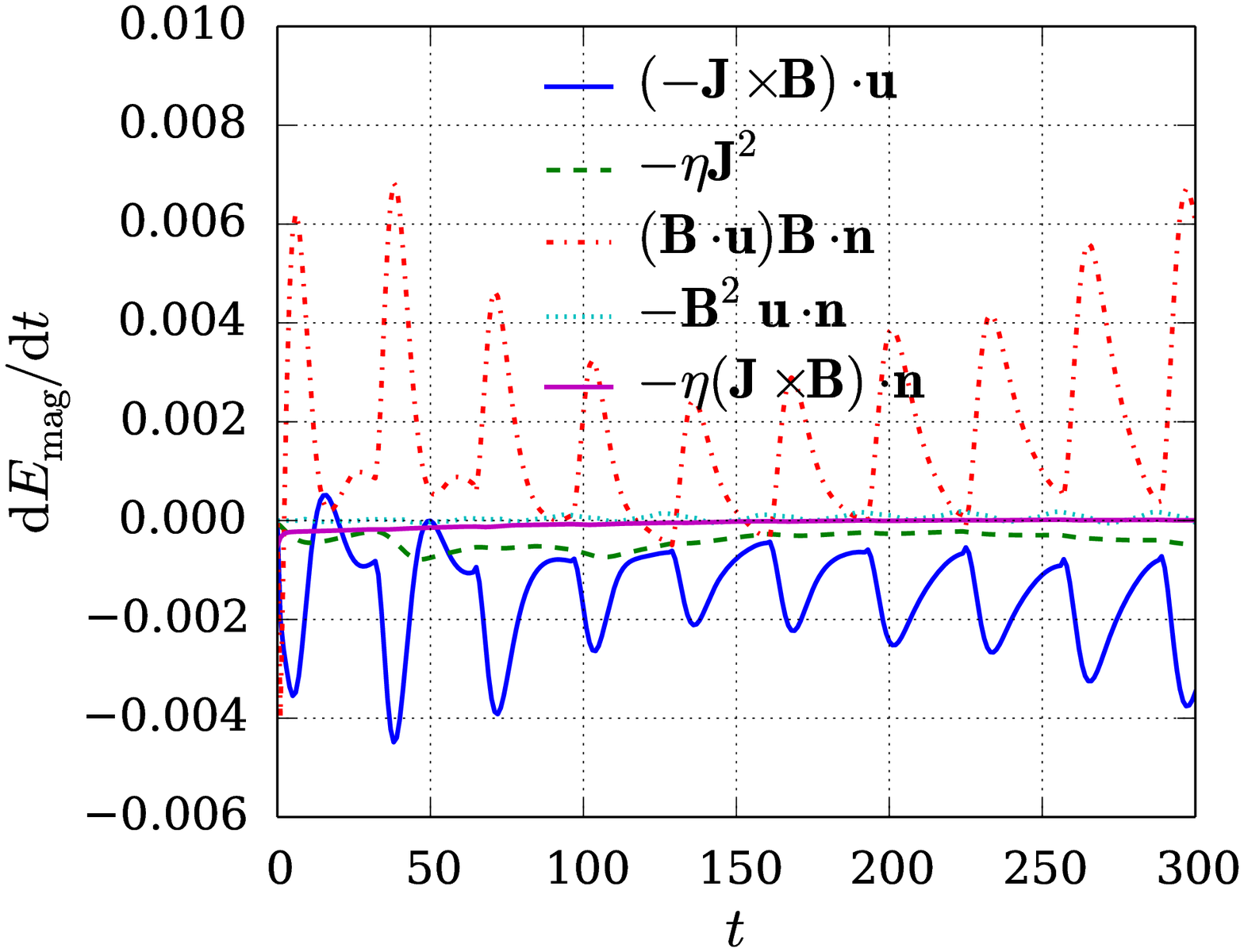} \\
\includegraphics[width=\columnwidth]{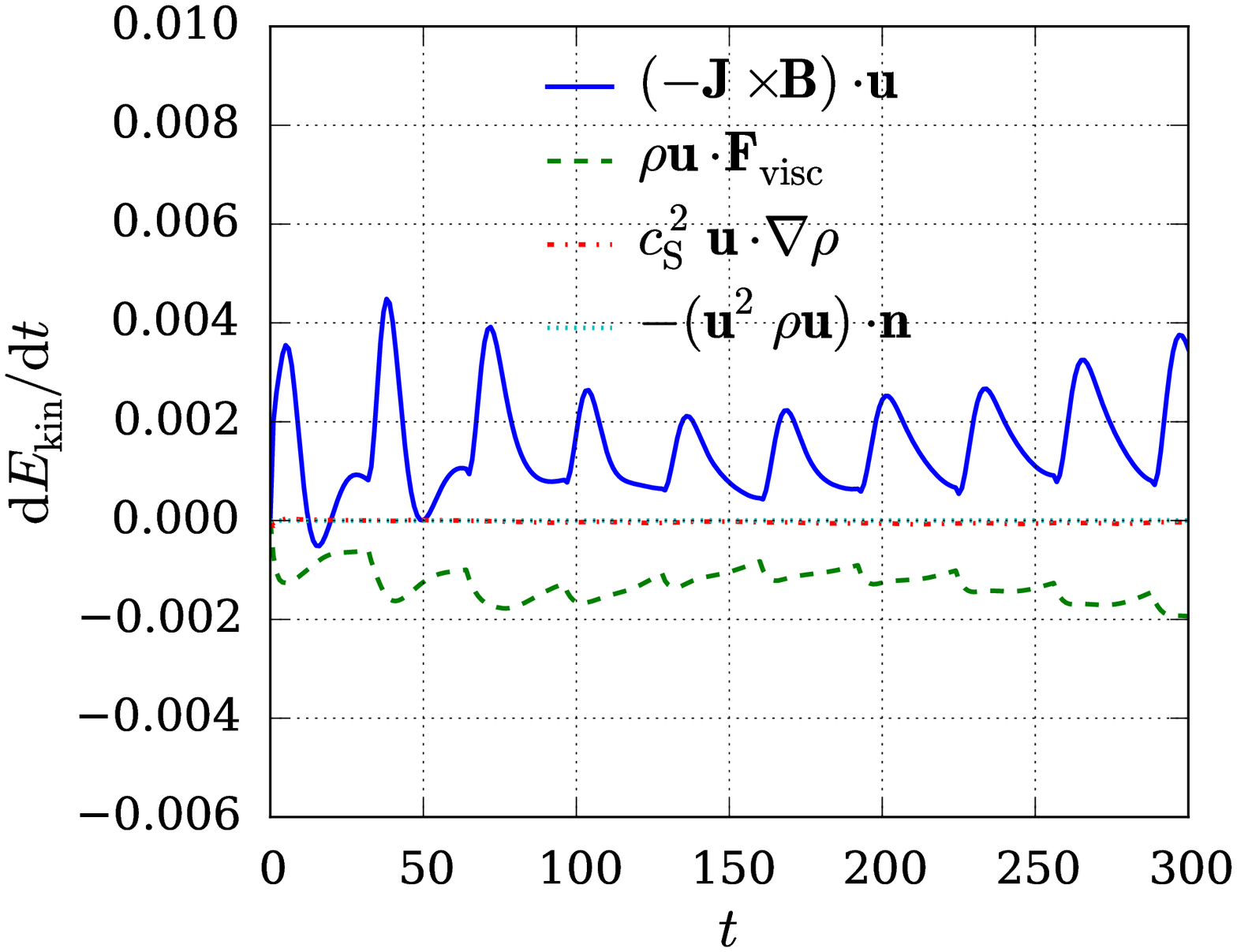}
\end{center}\caption[]{
Time evolution of the contributions to the magnetic energy change (upper panel)
and kinetic energy change (bottom panel) for the embedded parasitic polarity
initial condition.
Note that for the first two quantities in the list for the magnetic energy and the first
three quantities for the kinetic energy what we plot are the volume integrals, where
the volume is taken as the computational domain excluding the wave damping region.
Terms involving the normal vector $\nn$ are integrated over the surface of this volume.
}
\label{fig: dE_dt_para}
\end{figure}

It is clear from \Fig{fig: dE_dt_para} for the embedded polarities case
that magnetic energy is injected through the lower boundary.
From there it propagates into the domain where it is mostly converted into kinetic
energy through the Lorentz force.
At later times magnetic energy injection and conversion reach an
approximate equilibrium.

How is the magnetic energy dissipated?
Judging from the results (\Fig{fig: dE_dt_para}) the channel through
ohmic dissipation $\eta\JJ^2$ is rather limited due to the low value of $\eta$
compared to the energy input from the photosphere.
Similarly, other forms of magnetic energy fluxes are negligible
compared to the energy injection rate, like the Poynting flux through the upper
domain boundary and magnetic energy advection.
However, after conversion into kinetic energy, viscous effects are efficient enough
to account for a large part of the energy dissipation.

\begin{figure}[t!]\begin{center}
\includegraphics[width=\columnwidth]{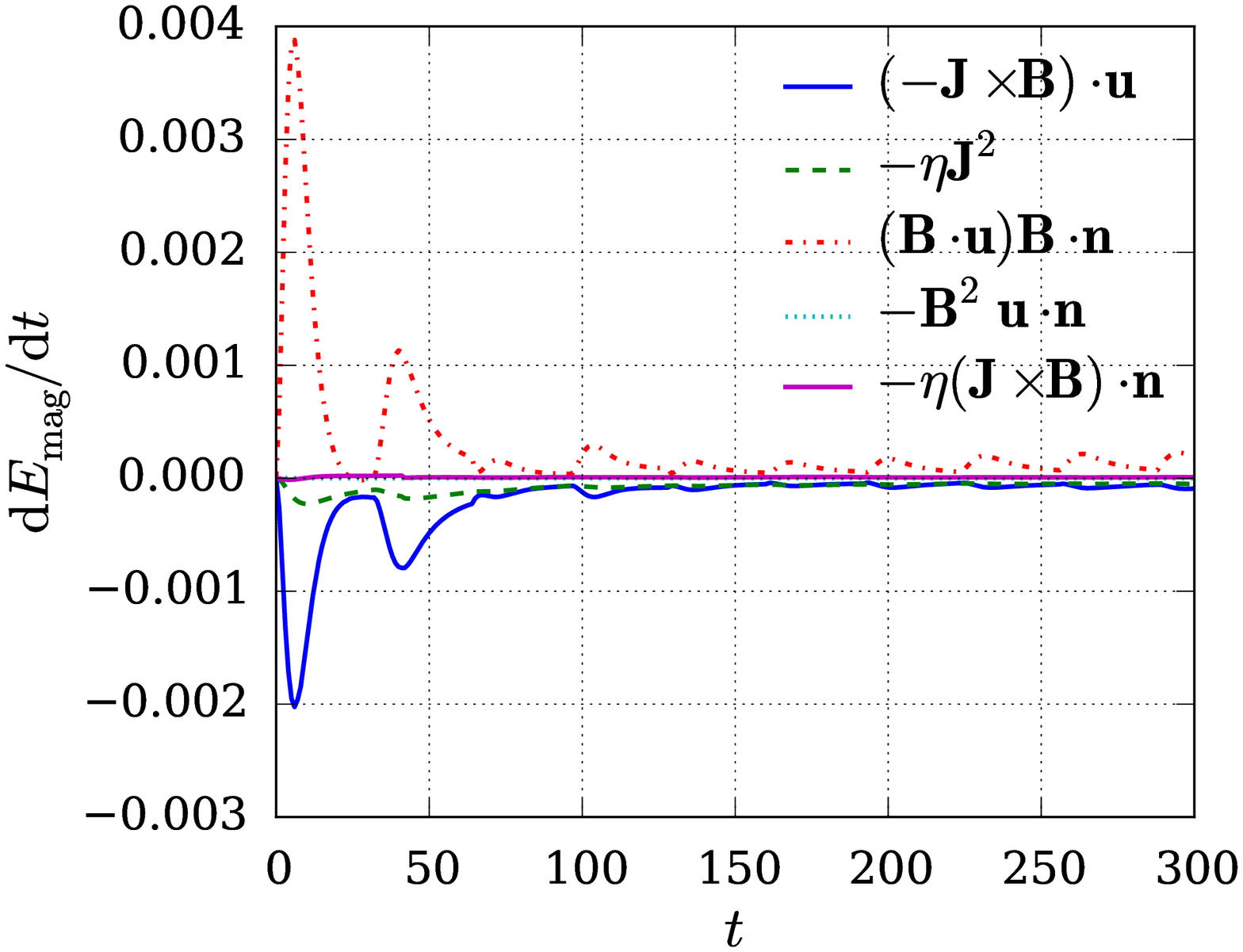} \\
\includegraphics[width=\columnwidth]{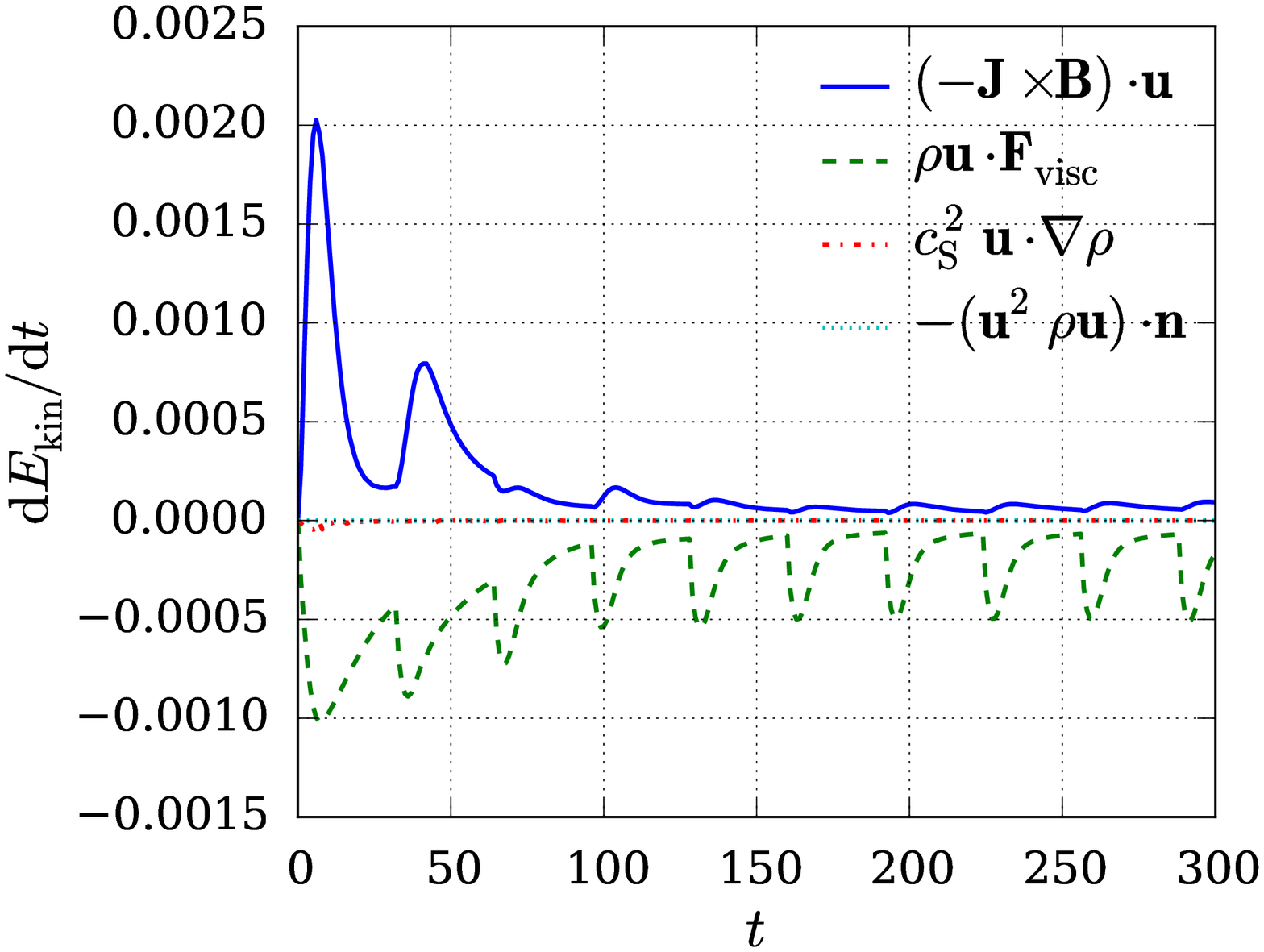}
\end{center}\caption[]{
Time evolution of the contributions to the magnetic energy change (upper panel)
and kinetic energy change (bottom panel) for the dominant polarity
initial condition.
Volume and surface integrals are taken as appropriate, as in \Fig{fig: dE_dt_para}.
}
\label{fig: dE_dt_dom}
\end{figure}

For the embedded dominant polarity case we first have to discount for effects
coming from the small non-potentiality of this configuration near the lower boundary.
To achieve this we perform a simulation without boundary driver and subtract
the values of the driven simulation from the values with $\uu_{\rm d} = 0$.
However, these effects become negligible after $t \approx 15$.
We then first observe a rather large release 
of magnetic energy, which is converted into kinetic energy (\Fig{fig: dE_dt_dom}).
This is due to the strong field close to the photosphere at $z = z_0$.
The conversion and dissipation channels are the same as for the embedded polarity
case, i.e.\ magnetic energy is converted into kinetic energy through the Lorentz
force and then mostly dissipated through viscous effects.
Over time we also observe a clear decrease for all terms, which is due to the
change of a strong near-surface field into a more homogeneous and weaker field
as the initial polarity regions are distorted and ultimately spread
out by a combination of the boundary flows and diffusion.

\subsection{Propagation of Energy to Higher Altitudes}\label{sec:propagation_to_altitude}
\begin{figure}[t!]\begin{center}
\includegraphics[width=\columnwidth]{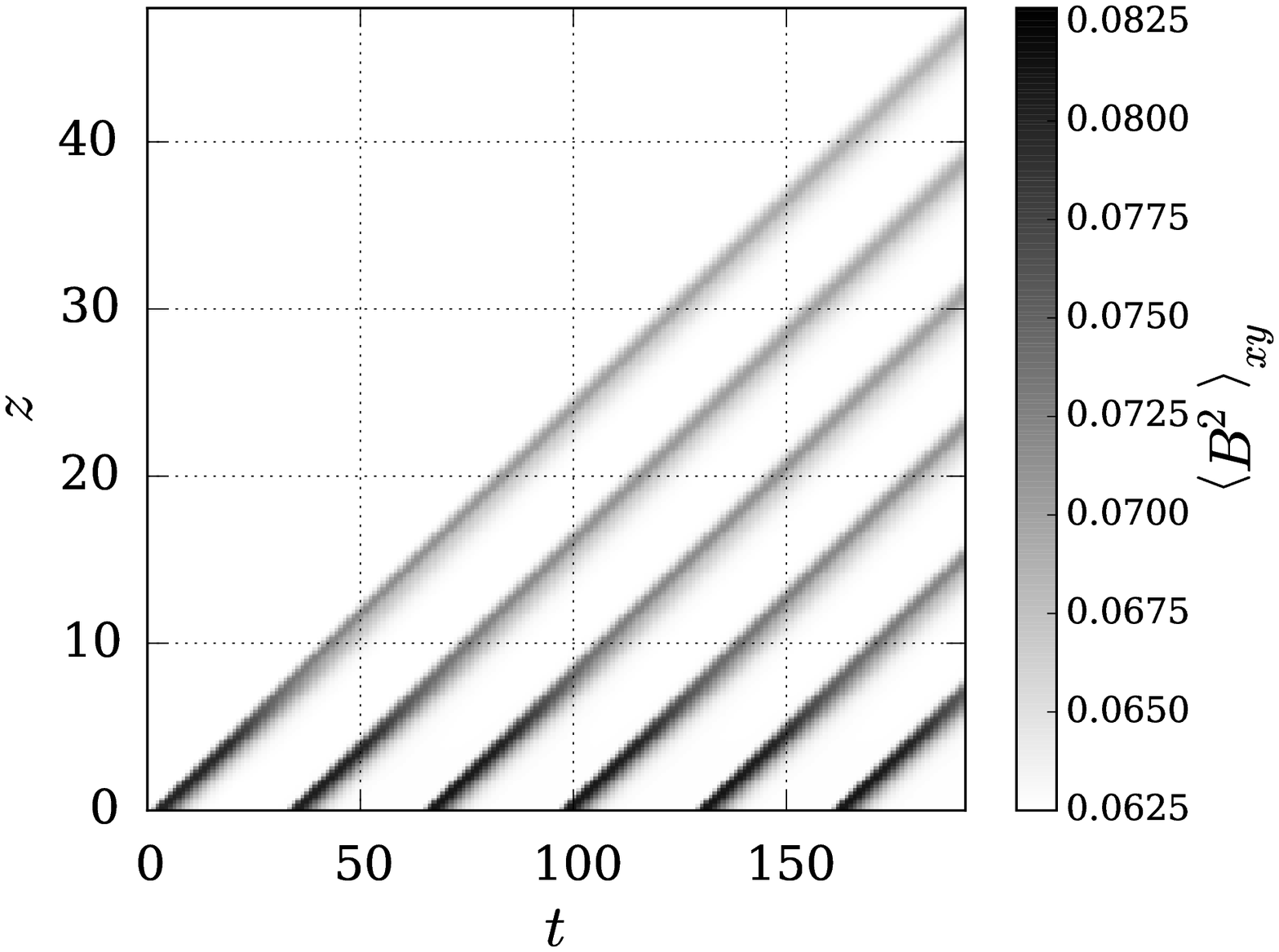} \\
\includegraphics[width=\columnwidth]{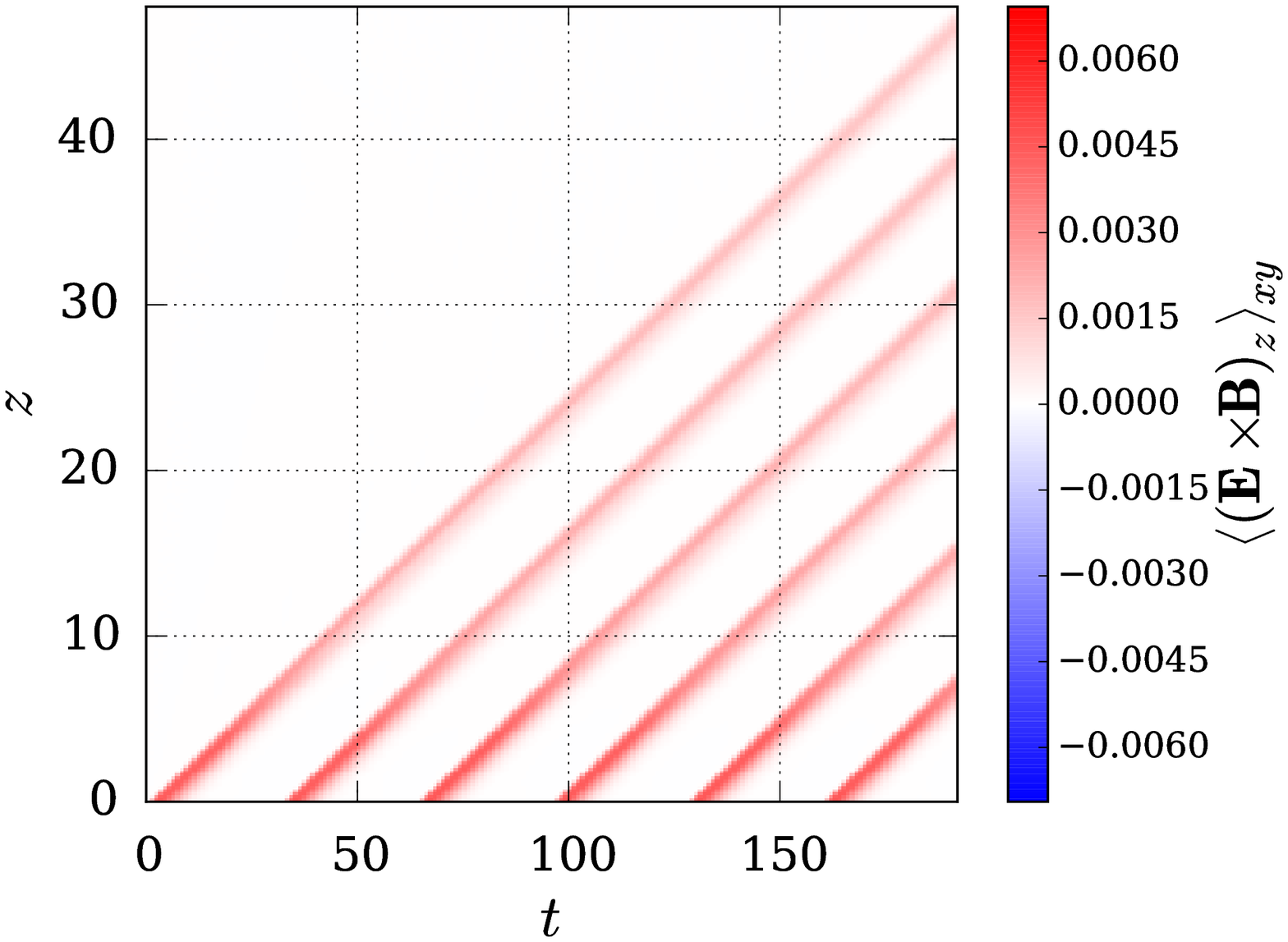}
\end{center}\caption[]{
Averages in the $xy$-plane of $B^2$ and the Poynting flux
in the $z$-direction $(\EE\times\BB)_z$ in dependence of the vertical
coordinate $z$ and time $t$ for the initially homogeneous case.
}
\label{fig: mz hom_e3}
\end{figure}

For the initially homogeneous field any energy or information is transported
through Alfv\'enic waves while acoustic waves appear to be insignificant.
Any magnetic field disturbance propagates freely into the domain
(\Fig{fig: mz hom_e3}) with the Poynting flux carrying the energy.
Due to the small value of the magnetic resistivity $\eta$ and viscosity $\nu$
the Alfv\'enic waves are only marginally damped, such that the energy
is efficiently transported to the top of the domain.
Since for this case the upper boundary allows for the reflection of
Alfv\'enic waves, we stop the simulation as soon as the first wave packet
reaches the boundary.

\begin{figure}[t!]\begin{center}
\includegraphics[width=\columnwidth]{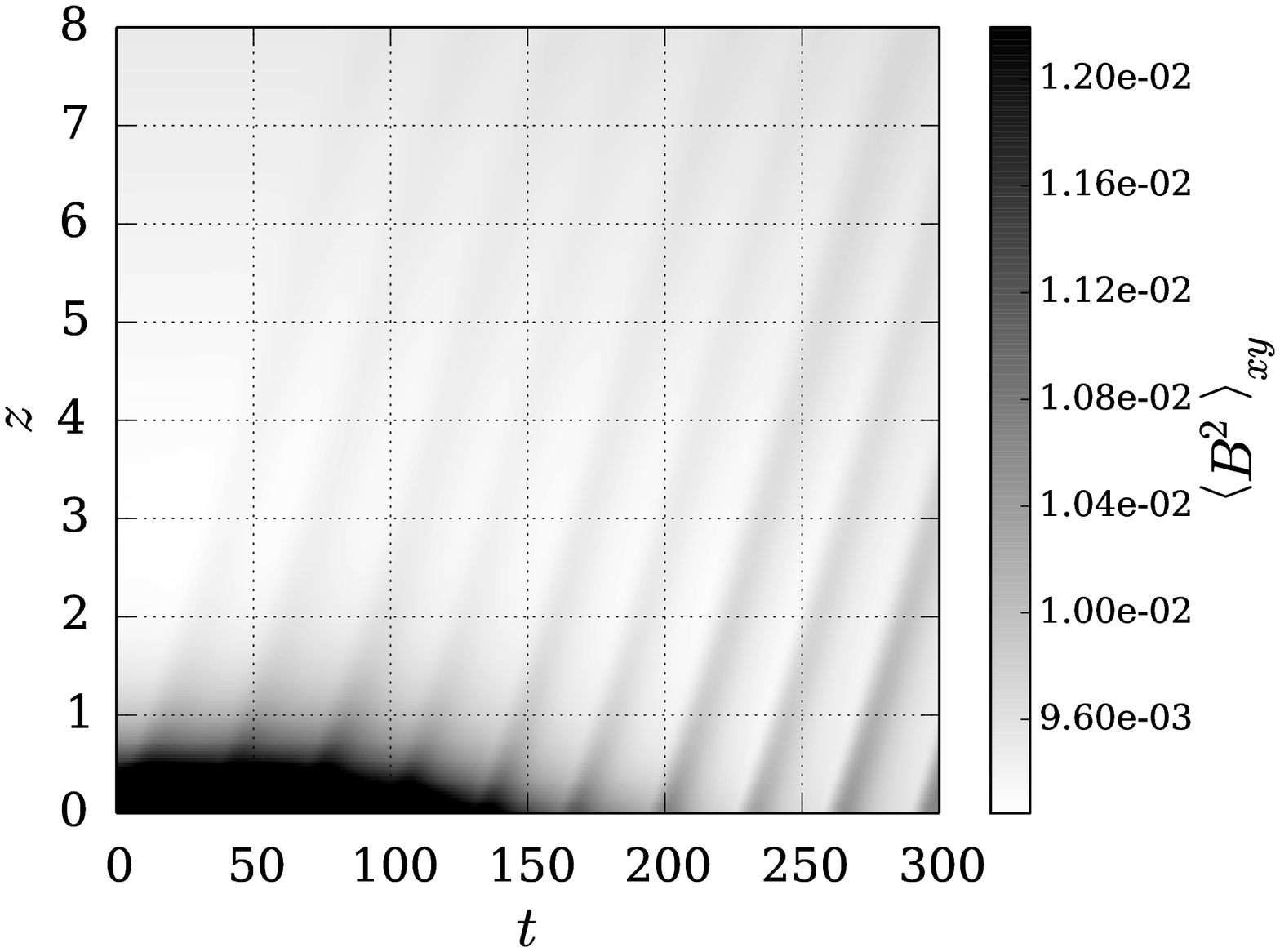} \\
\includegraphics[width=\columnwidth]{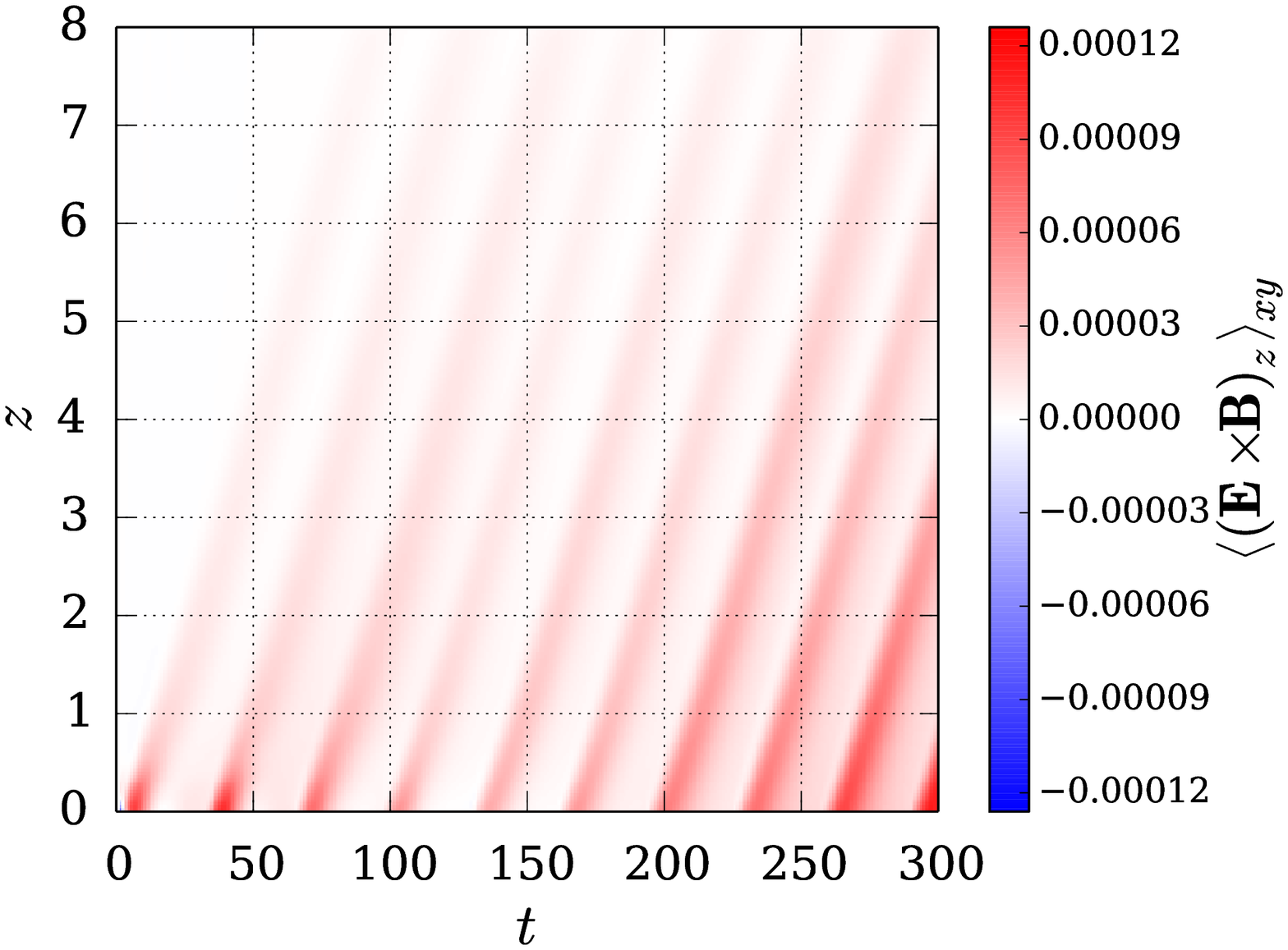}
\end{center}\caption[]{
Averages in the $xy$-plane of $B^2$ and the Poynting flux
in the $z$-direction $(\EE\times\BB)_z$ in dependence of the vertical
coordinate $z$ and time $t$ for the parasitic polarity case.
}
\label{fig: mz para_e3}
\end{figure}

For the parasitic polarity configuration the energy from the footpoint
motion is initially trapped at low heights (\Fig{fig: mz para_e3}), primarily below the null-points and 
the separatrix domes.
This is due to the trapping of Alfv\'enic waves through closed (to the photosphere)
magnetic field lines.
For $100\leq t \leq 200$, we observe a restructuring of the
magnetic skeleton, as described in Section \ref{sec:skeleton},
which is characterized by a shrinking of the domes as the 
parasitic polarities are `shredded' by the photospheric flows.
This leads to a flux of magnetic null points through the lower boundary and their
subsequent disappearance (\Fig{fig: nulls}).
As a result, there is now a larger fraction of the 
field lines at the photosphere that are open
allowing the injected twist to travel into the domain.
Note that the increase in efficiency of energy propagation to large heights coincides with the disappearance of
the magnetic null points at time $t \approx 210$.

\begin{figure}[t!]\begin{center}
\includegraphics[width=\columnwidth]{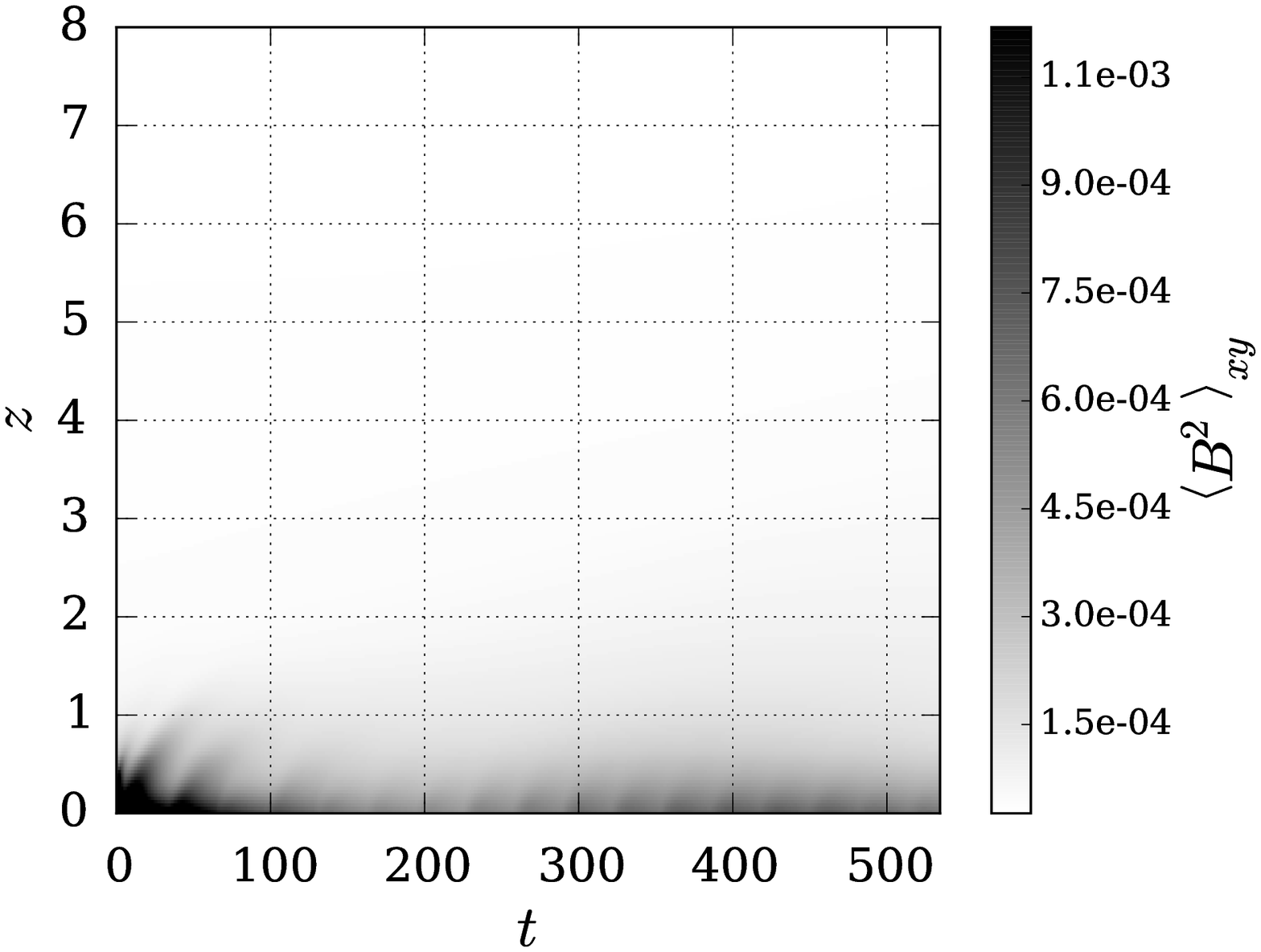} \\
\includegraphics[width=\columnwidth]{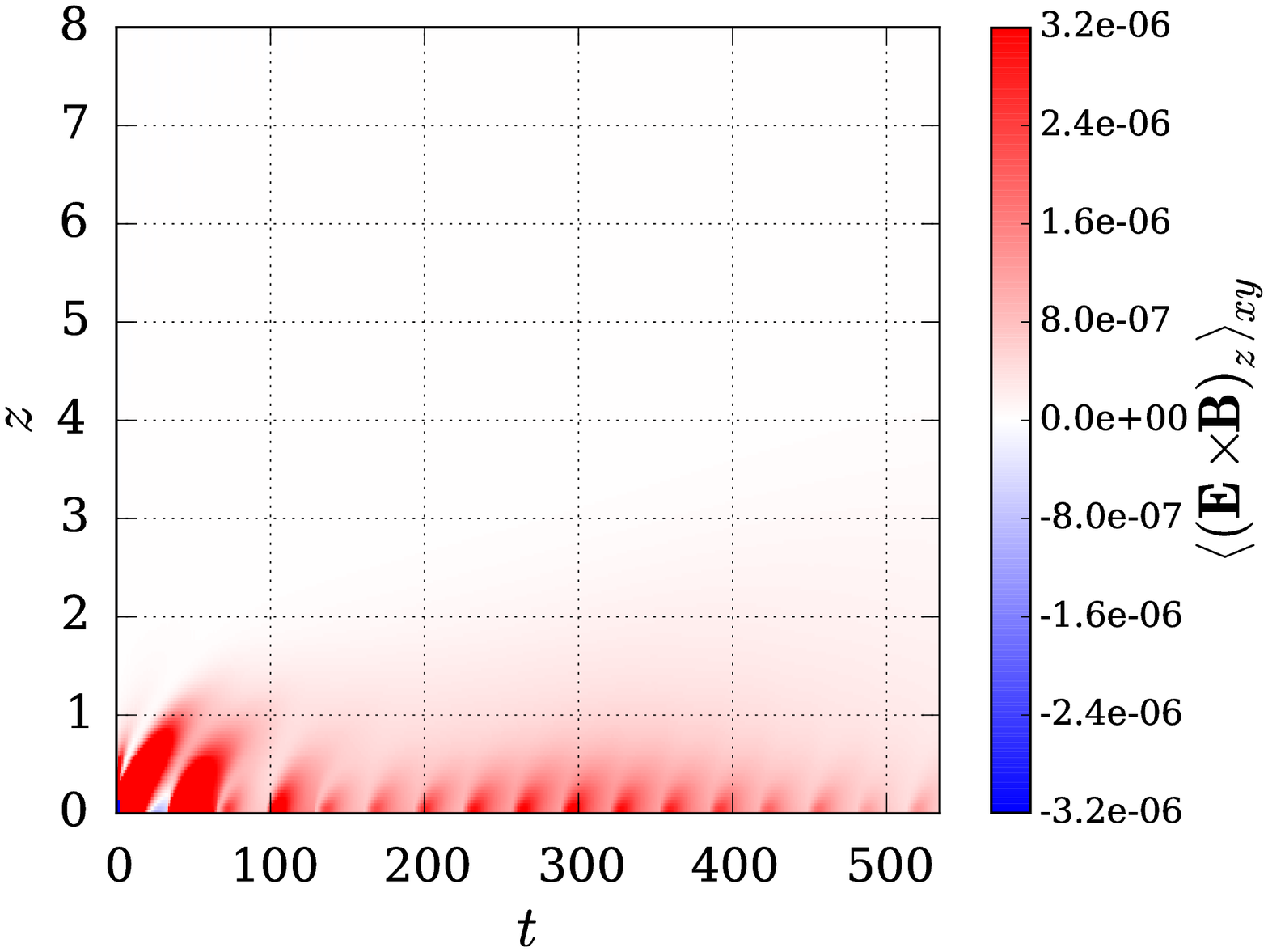}
\end{center}\caption[]{
Averages in the $xy$-plane of $B^2$ and the Poynting flux
in the $z$-direction $(\EE\times\BB)_z$ in dependence of the vertical
coordinate $z$ and time $t$ for the dominant polarities case.
}
\label{fig: mz dom_e3}
\end{figure}

Similarly, for the dominant polarity case we observe a trapping of 
magnetic energy below the locations of the magnetic nulls.
Due to the strong field, most of the magnetic energy is stored close to the photosphere.
This holds until the breakage of the field's topology into a simpler structure
which allows for fluxes into the domain.
Therefore, we observe energy fluxes after time $t\approx 200$ which
more easily reach the top boundary (\Fig{fig: mz dom_e3}).
We quantify the efficiency of the propagation via the ratio of the Poynting flux
at $z = 2$ to the value at $z = 0$.
Since the Alfv\'en speed varies with height we take the values at $z = 2$ with a time
delay of $100$ code time units which gives us a reasonably good estimate.
By doing so we find a ratio for the Poynting flux of ca. $0.35\%$ for waves
emitted at $t = 64$ and a ratio of $2.8\%$ for $t = 415$.
This shows that the energy flux is enhanced after the break up of the magnetic field topology.

\section{Discussion and Conclusions}
We have considered above the application of boundary flows to three different model 
coronal fields. In the first, most simplified model an initially homogeneous field was 
used.
In this case we showed that it is feasible to induce braiding to the magnetic
field of the solar corona by motions on the photosphere.
Moving to the more realistic
models with a mixed polarity photospheric field, the energy transport to large 
altitudes was inhibited by the complex field topology.
In this work we did not include a stratified atmosphere in
which the Alfv\'en speed can change by several orders of magnitude.
It was shown by \cite{Ballegooijen-Asgari-Targhi-2014-787-87-ApJ} that this has
a strong effect on the propagation and dissipation of energy, and should therefore be considered in a future study.
While for a large Alfv\'en speeds, compared to the driving velocities, the DC
heating dominates, for small Alfv\'en speeds AC dominates.
Furthermore, non-linear effects lead to the dissipation of counter propagating waves.

\begin{figure}[t!]\begin{center}
\includegraphics[width=0.9\columnwidth]{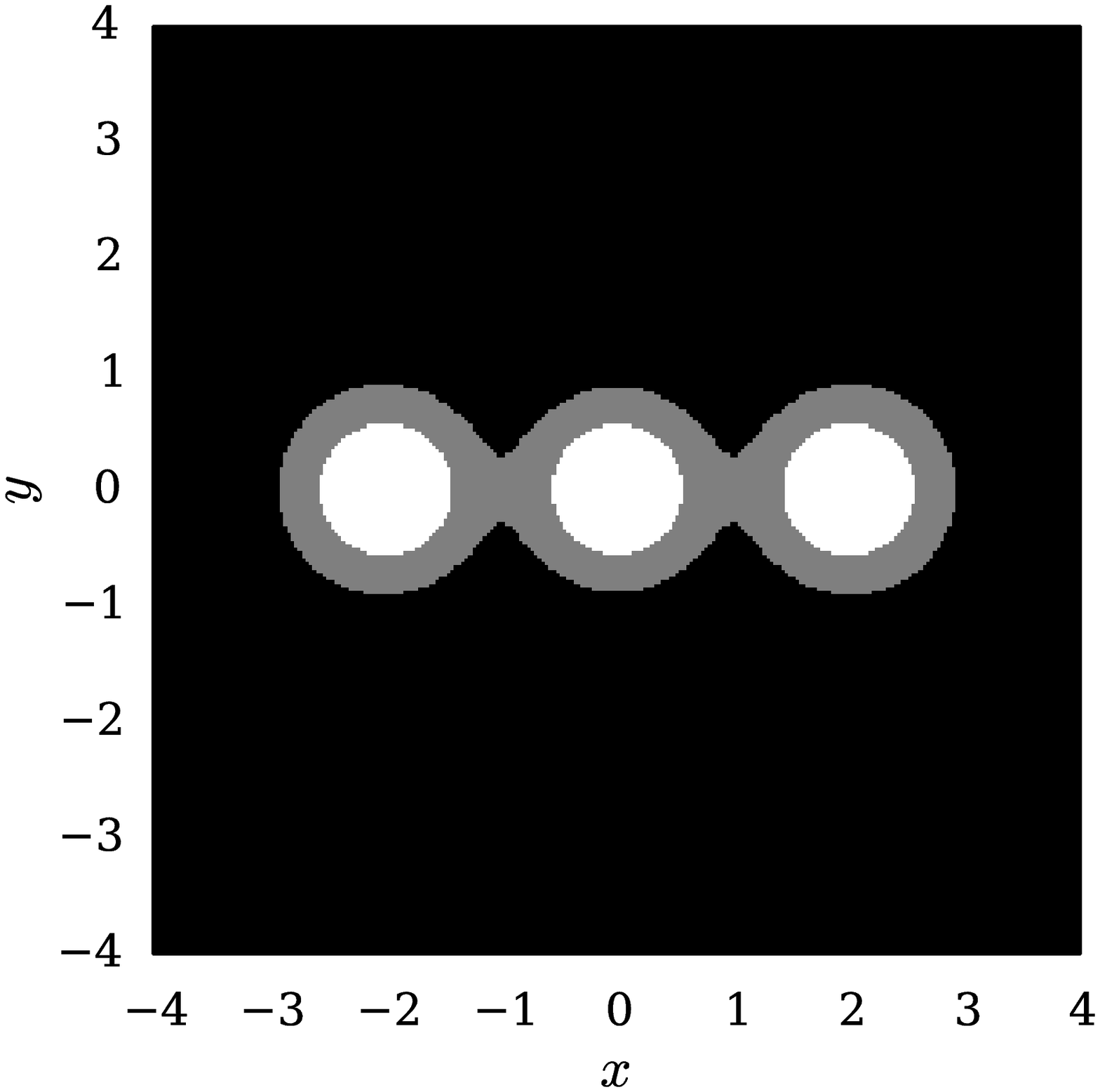} \\
\includegraphics[width=0.9\columnwidth]{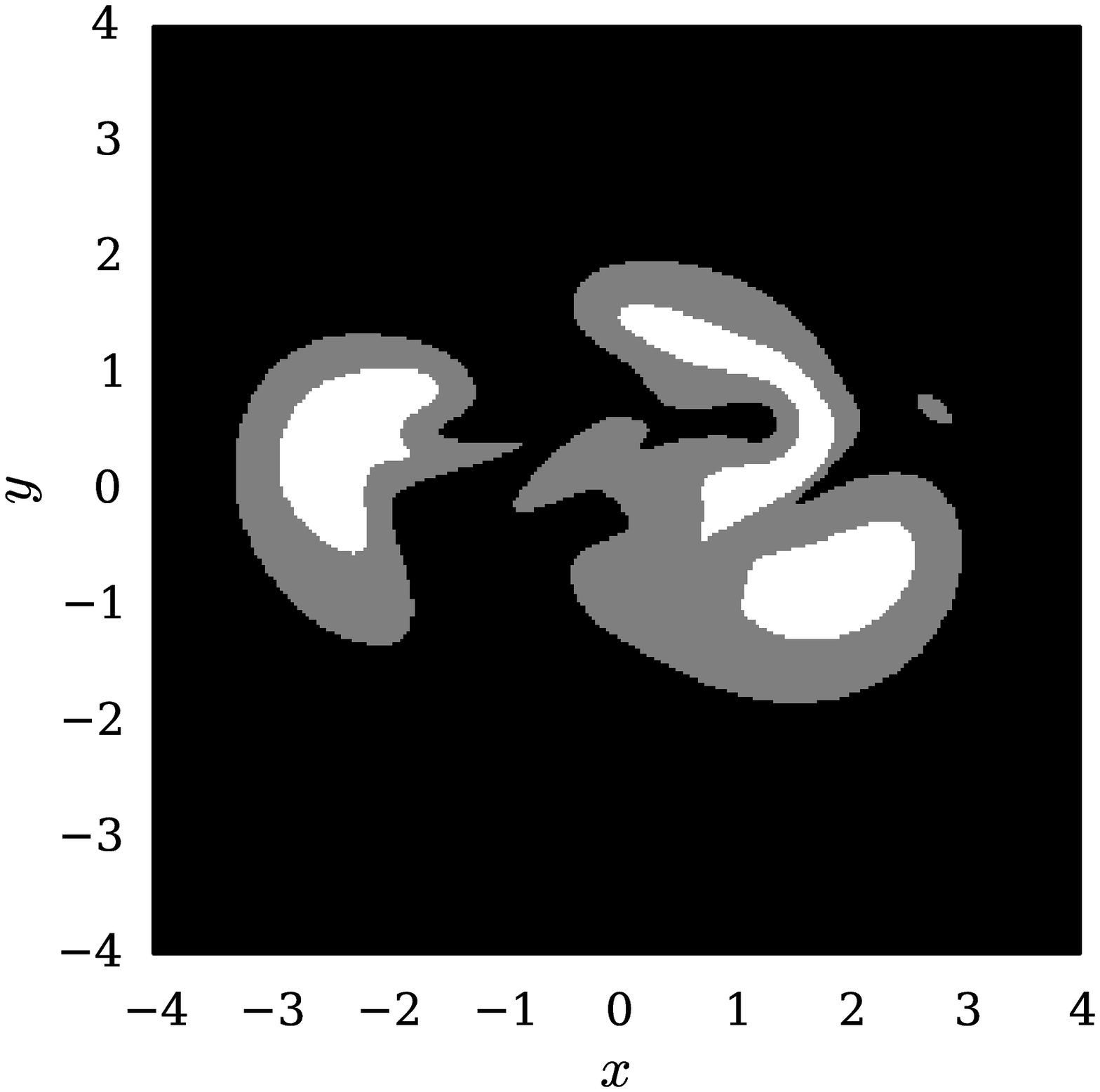} \\
\includegraphics[width=0.9\columnwidth]{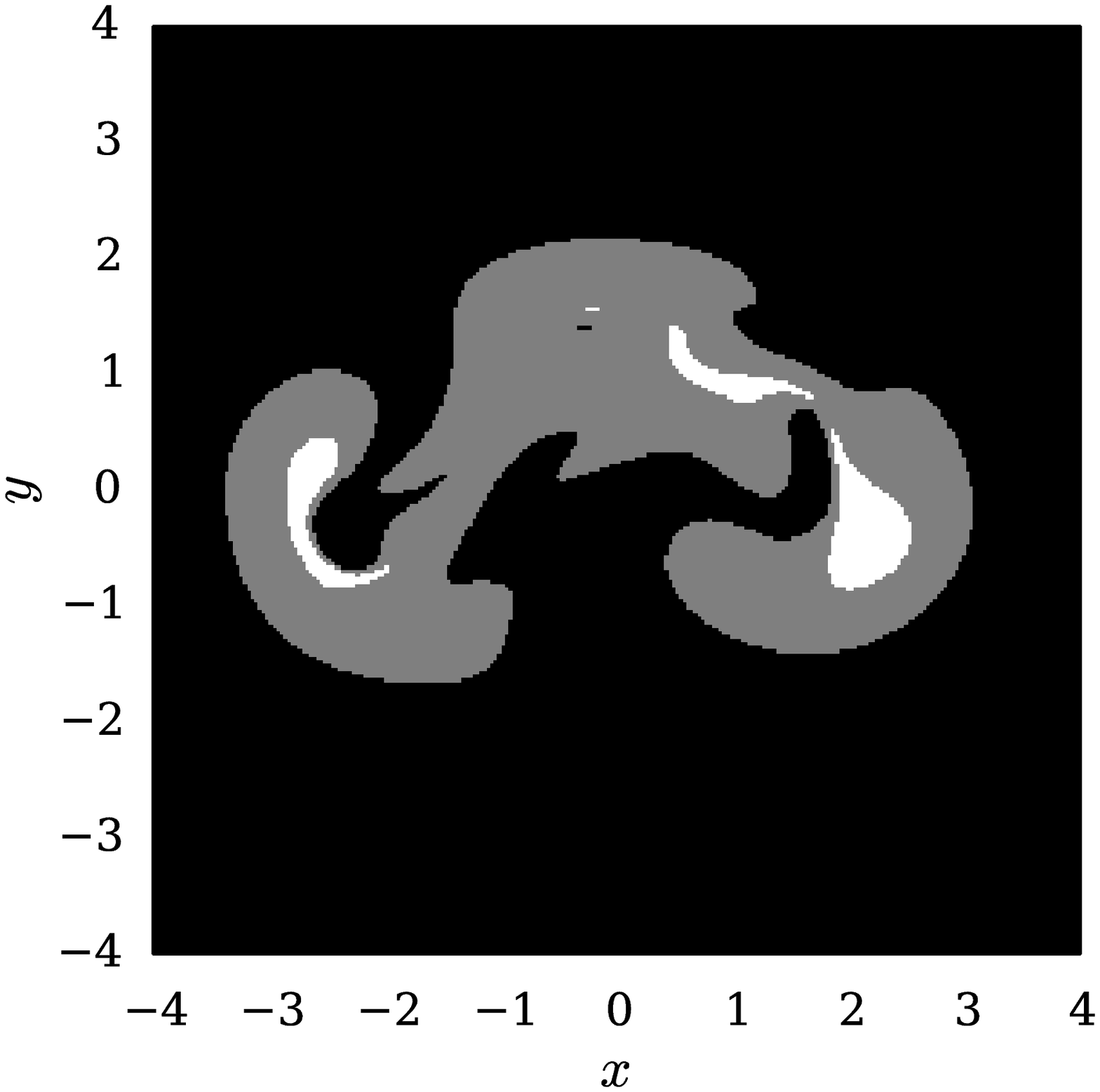}
\end{center}\caption[]{
Synthetic magnetogram for the parasitic polarity case at times $t = 0$ (top panel),
$t = 58$ (central panel) and $t = 109$ (bottom panel).
The shadings correspond to the reduced $z$-component of the magnetic field with black
positive and white negative polarity.
}
\label{fig: bbz para}
\end{figure}

\begin{figure}[t!]\begin{center}
\includegraphics[width=0.9\columnwidth]{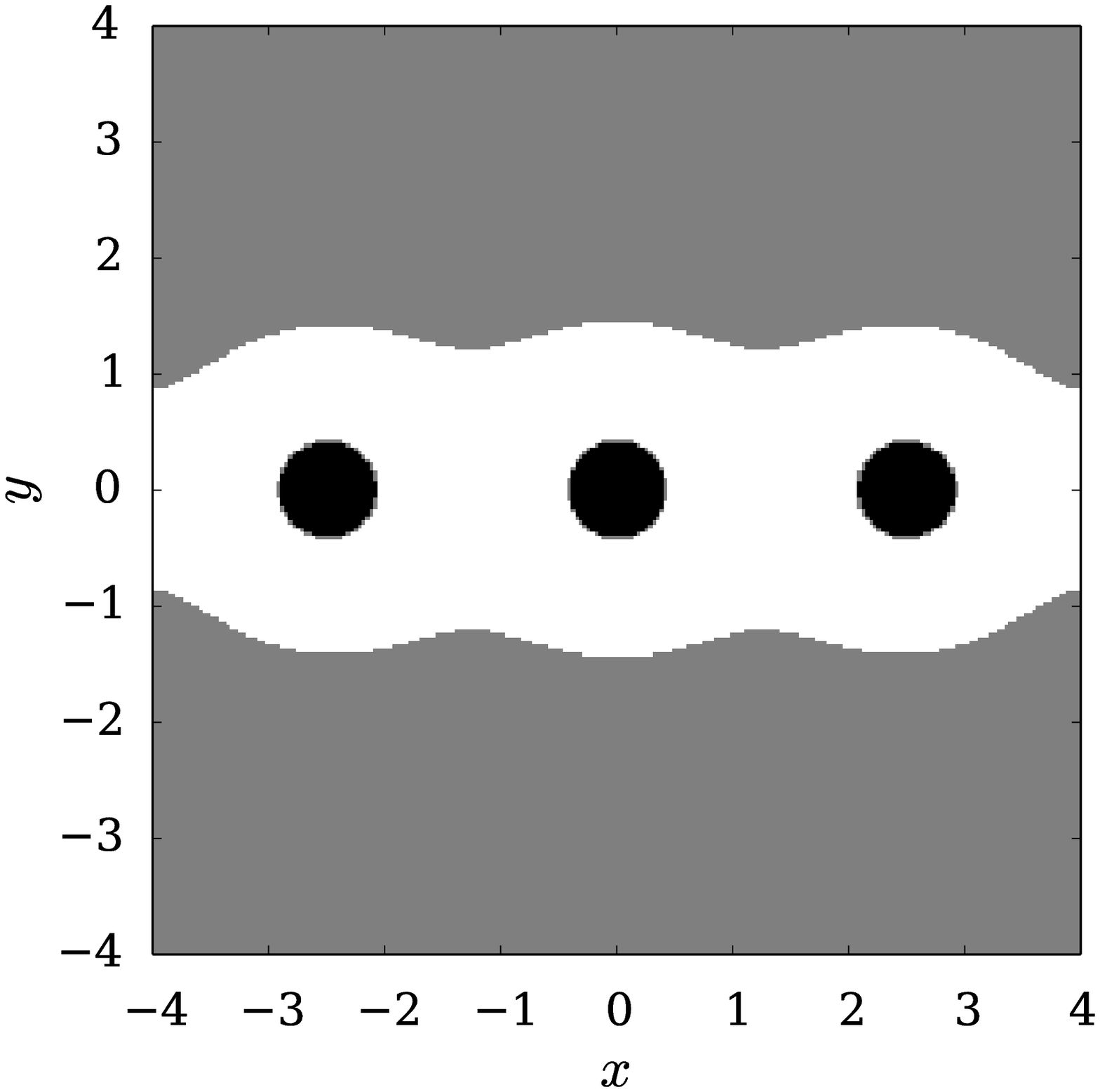} \\
\includegraphics[width=0.9\columnwidth]{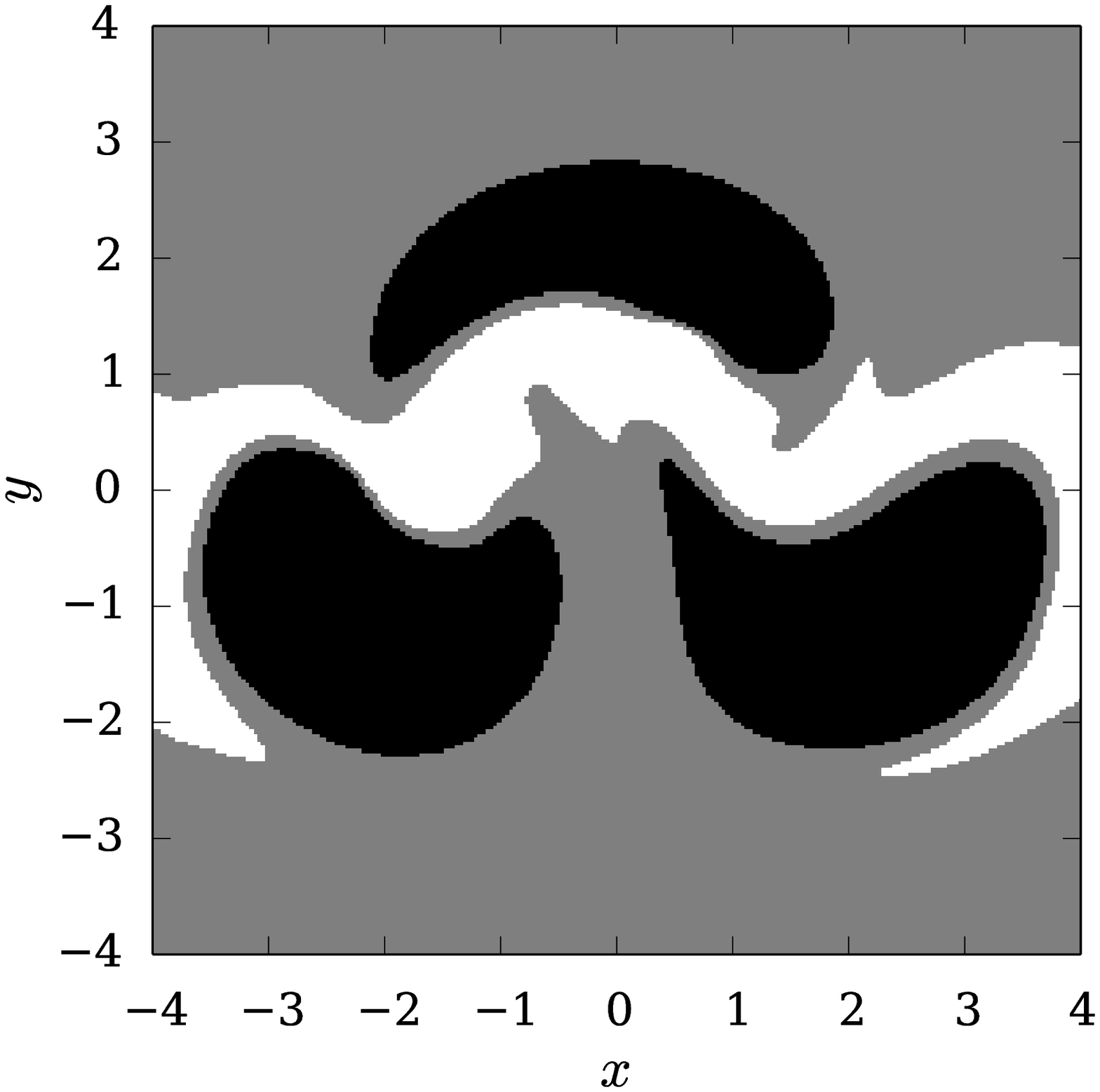} \\
\includegraphics[width=0.9\columnwidth]{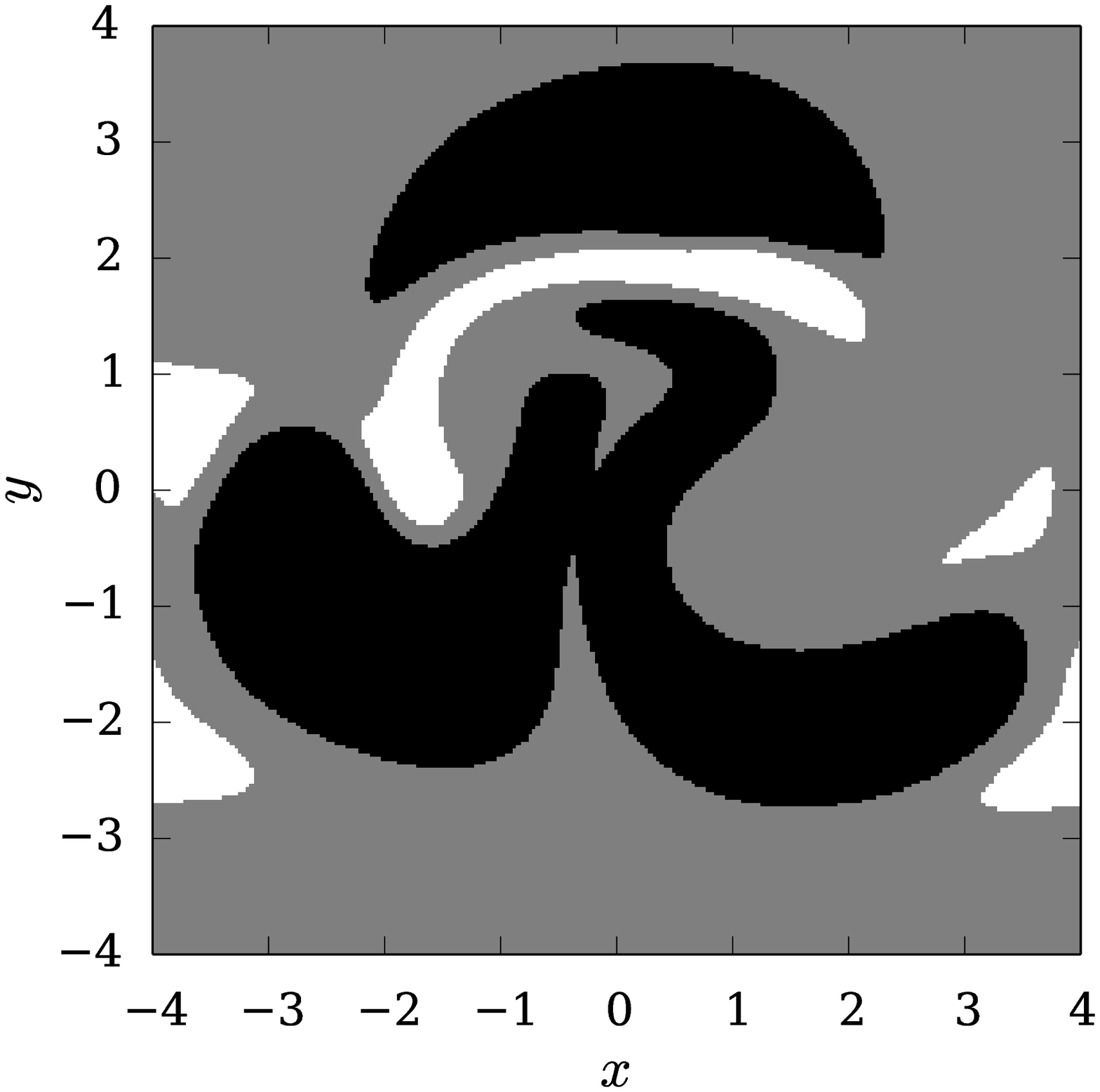}
\end{center}\caption[]{
Synthetic magnetogram for the dominant polarity case at times $t = 0$ (top panel)
$t = 95$ (central panel) and $t = 185$ (bottom panel).
The shadings correspond to the reduced $z$-component of the magnetic field with black
positive and white negative polarity.
}
\label{fig: bbz dom}
\end{figure}

In order to understand our results in the context of the corona
we can extract synthetic magnetograms of the line-of-sight magnetic field
from the magnetic field on the lower boundary of our simulation domain, and compare with 
processes occurring  in observed solar magnetograms.
Here we take the line of sight to be simply the $z$-direction.
In order to compare to actual observations we reduce the $z$-component of the magnetic
field to three values.
Specifically, it is set to $+1$ at points where $B_z(z = 0) > B_{\rm cut}$, 
$-1$ if $B_z(z = 0) < -B_{\rm cut}$ and $0$ otherwise 
(to simulate the noise threshold on magnetogram observations).
For the parasitic polarity case, shown in \Fig{fig: bbz para}, we choose $B_{\rm cut} = 0.15\times |B_z(x, y, z = 0, t)|_{\rm max}$
and for the dominant polarity case $B_{\rm cut} = 0.003\times |B_z(x, y, z = 0, t)|_{\rm max}$ (\Fig{fig: bbz dom}).
Note that we take the maximum over $x$, $y$ and $t$, which means that the cut off value
is fixed in time.

From the synthetic magnetograms for both simulations we clearly
observe a complex interaction of opposite polarity regions which may lead to both the 
splitting and merging of polarity regions (sometime called ``flux fragments").
At later times, however, due to the overall mixing and cancellation of the polarities
(a result of both the stretching nature of the imposed flow and diffusion) 
we end up with one surviving polarity region (positive).
This behaviour is consistent
with the observed behaviour of magnetic flux fragments on the Sun, which are 
known to undergo a continuous process of merging and fragmentation 
\citep[e.g.][and references therein]{deforest2007}.
It is important to note 
that this `shredding' of the magnetic flux fragments in the synthetic magnetograms
is crucially dependent on the relative length scales of the flows and the flux 
fragments; in our case the flows have significant gradients over the scale
of the initial fragments. 
From this point of view, our results are probably best compared to local regions of
the photosphere in which the flux fragments are comparable to
the scale of the underlying motions \citep{Gosic-Rubio-2014-797-49-ApJ}.

In the two models with embedded parasitic/dominant polarities,
we investigated the effects of footpoint motions on fields where
a significant fraction of the field lines initially connect back to the 
photosphere, while others connect to the upper boundary.
The applied photospheric motions do not create new magnetic flux, but, as discussed above,
induce shredding of existing flux, leading in the magnetograms to the 
eventual `death' of the flux fragments \citep{lamb2013}.
On the Sun, this process is in
a statistically steady state with the competing process of emergence of new flux -- 
that we exclude from our simplified model.

We showed that in the initial stages of the simulations with mixed polarity, 
the presence of closed field lines restricted the energy propagation
into the domain.
However, as we continue with the driving, the embedded polarities are shredded 
into weaker fragments.
This reduces the range over which they influence the coronal field.
In particular, we have shown that it results in a reduction in the number, and perhaps more importantly the height of the coronal null points.
This is particularly clear in the case of the embedded parasitic polarities, 
where the separatrix domes enclosing the closed flux get progressively smaller (on 
average, both in height and in extent over the $xy$-plane) as the simulation proceeds.
As a result, the propagation of disturbances can access the open field regions more 
readily, and the energy is propagated much more efficiently to large heights.

In summary, we can confirm the feasibility of energy and disturbance
propagation from the photosphere into the corona through the motion
of footpoints.
The magnetic field topology plays an essential role during this process
with a magnetic carpet structure containing nulls largely inhibiting the process.
We showed that the shredding of magnetic polarities
by photospheric flows leads to a simplification of the 
magnetic topology through the disappearance of null points
(either through the lower boundary or in annihilation processes 
within the volume).
This in turn allows a more effective propagation of
energy to large heights in the corona.

\acknowledgments

All the authors acknowledge financial support from the UK's
STFC (grant number ST/K000993).
We also thank the anonymous referee for the useful comments which helped improve this paper.

\bibliography{references}

\end{document}

%% file: macros.tex
\newcommand{\EQ}{\begin{equation}}
\newcommand{\EN}{\end{equation}}
\newcommand{\EQA}{\begin{eqnarray}}
\newcommand{\ENA}{\end{eqnarray}}

\newcommand{\Eqss}[2]{Equations~(\ref{#1})--(\ref{#2})}

\newcommand{\Fig}[1]{Figure~\ref{#1}}

{}
{}
{}

{}
{}
{}
{}
{}
{}
{}
{}
{}
{}
{}
{}
{}
{}
{}
{}
{}
{}
{}
{}
{}

{}
{}
{}

{}

{}
{}

\newcommand{\eee}{\hat{\mbox{\boldmath $e$}} {}}

\newcommand{\xx}{\bm{x}}

\newcommand{\BB}{\bm{B}}

\newcommand{\uu}{\mbox{\boldmath $u$} {}}

\newcommand{\EE}{\mbox{\boldmath $E$} {}}

\newcommand{\JJ}{\mbox{\boldmath $J$} {}}
\newcommand{\SSS}{\mbox{\boldmath $S$} {}}
\newcommand{\AAA}{\mbox{\boldmath $A$} {}}

\newcommand{\ee}{\mbox{\boldmath $e$} {}}
\newcommand{\nn}{\mbox{\boldmath $n$} {}}

\newcommand{\FF}{\mbox{\boldmath $F$} {}}

\newcommand{\nab}{\mbox{\boldmath $\nabla$} {}}

\newcommand{\DD}{{\rm D} {}}

\newcommand{\dd}{{\rm d} {}}

\def\cs{c_{\rm s}}